\documentclass[12pt]{article}

\usepackage{amssymb,amsmath}
\usepackage[dvips]{graphicx}
\usepackage{harvard}
\usepackage{amssymb}

\def\onehalf{{\scriptstyle\frac{1}{2}}}

\let\ro=\varrho

\def\R{\mathbb R}

\newtheorem{theorem}{Theorem}[section]
\newtheorem{definition}{Definition}[section]
\newtheorem{lemma}{Lemma}[section]
\newtheorem{corollary}{Corollary}[section]

\newtheorem{remark}{Remark}[section]

\title{Early exercise boundary for American type of floating strike Asian option and its numerical approximation}

\author{Tom\'a\v{s} Bokes and  Daniel \v{S}ev\v{c}ovi\v{c} \\
{\normalsize Department of Applied Mathematics and Statistics,}\\ 
{\normalsize Faculty of Mathematics, Physics and Informatics,} \\
{\normalsize Comenius University, 842 48 Bratislava, Slovak Republic}}

\begin{document}

%\AlgLogo{1}{10}

\maketitle

\begin{abstract}
In this paper we generalize and analyze the model for pricing American-style Asian options due to \cite{ATH-PLJ-2000} by including a continuous dividend rate $q$ and a general method of averaging of the floating strike. We focus on the qualitative and quantitative analysis of the early exercise boundary. The first order Taylor series expansion of the early exercise boundary close to expiry is constructed. We furthermore propose an efficient numerical algorithm for determining the early exercise boundary position based on the front fixing method. Construction of the algorithm is based on a solution to a nonlocal parabolic partial differential equation for the transformed variable representing the synthesized portfolio. Various numerical results and comparisons of our numerical method and the method developed by \cite{MD-YKK-2006} are presented.


\noindent {\bf AMS-MOS classification:} 35K15, 35K55, 90A09, 91B28

\end{abstract}

\pagestyle{myheadings}
\thispagestyle{plain}
\markboth{T. Bokes \& D. \v{S}ev\v{c}ovi\v{c} }{Early exercise boundary for American type of Asian option}

\markboth{}{}

\section{Introduction}
Evolution of trading systems influences the development of financial derivatives market. First, simple derivatives like forwards and plain vanilla options were used to hedge the risk of a portfolio. Progress in pricing these simple financial instruments pushed traders into inventing less predictable and more complex derivatives. Using financial derivatives with more complicated pay-offs brings into attention also new mathematical problems. Asian options belong to a group of the so-called path-dependent options. Their pay-off diagrams depend on the spot value of the underlying during the whole or some part(s) of the life span of the option. Usually Asian options depend on the (arithmetic or geometric) average of the spot price of the underlying. They can be also used as a useful tool for hedging highly volatile assets or goods. Since the price of an underlying varies during the life span of the option the holder of the Asian option can be secured for the risk a sudden price jumping to an undesirable region (too high for the call option holder or too low for the put option holder). Among path dependent options Asian options plays an important role as they are quite common in currency and commodity markets like e.g. oil industry (cf. \cite{PW-SH-JD-1995,JCH-1997,LW-YKK-HY-1999,ATH-PLJ-2000,JD-2006,MD-YKK-2006,UW-2006,YKK-2008,BCK-YSO-2004,RW-MCF-2003,VL-2004}).
	
In this paper we focus on the so-called floating strike Asian call or put options whose strike price depends on the averaged path history of the underlying asset. More precisely, we are interested in pricing American-style Asian call and put options having the pay-off functions $V_T(S,A)=(S-A)^+$ and $V_T(S,A)=(A-S)^+$, resp.
The strike price $A$ is given as an average of the underlying over the time history $[0,T]$. We are analyzing the early exercise boundary for American-style Asian option (cf. \cite{ATH-PLJ-2000,MD-YKK-2006,YKK-2008}). Recall that American-style options can be exercised at any time until the maturity $T$. The holder of such an option has the right to exercise it or to keep it depending on the spot price of the underlying $S_t$ at time $t$ and its history $\{S_u, 0\le u\le t\}$ prior the time $t$. The boundary between ''continuation'' and ''stopping'' regions plays an important role in pricing American-style of options. It can be described by the mapping $S^*_t: t\mapsto S^*_t$, where $S^*_t$ is the so-called early exercise boundary (cf. \cite{JCH-1997,RG-HEJ-1984,RG-RR-1984,IK-1988,JC-2008,YKK-2008,RAK-JBK-1998,RM-2002,AP-2008}.

The paper is organized as follows. In the next section, we discuss the probabilistic model for valuation of  American type Asian options with a floating strike given in the form of an average of the underlying asset price.  The model is based on conditioned expected values and theory of martingales. In the third section, we present a general result enabling us to calculate the value of the limit of the early exercise boundary at expiry. We also calculate the analytical integral formula for an option with continuous geometric average and an approximation for the value of an option with a continuous arithmetic average. The main result of this section is the approximation formula for the first order Taylor series approximation of the early exercise boundary near to expiry. Similarly as in the case of plain vanilla call options, we show that the leading order of the expansion is the square root $\sqrt{T-t}$ of the time $T-t$ remaining to maturity. In the fourth section, we present the fixed domain transformation method yielding a nonlocal nonlinear parabolic partial differential equation for pricing the synthesized portfolio for an Asian option. We also present an efficient and robust numerical scheme for construction of an approximation of the solution to the governing system consisting of a partial differential equation with an algebraic constraint. In the last part of this section we present also the results of the presented method on some examples.

\section{A probabilistic approach for pricing of American-Style of Asian options}

The main purpose of this section is to derive an integral equation for valuation of the early exercise boundary of an American-style Asian option paying continuous dividends. We follow the ideas of derivation due to \cite{ATH-PLJ-2000}. Their formula for a floating strike option was derived using the theory of martingales and conditioned expected values. We extend their formula to Asian options on underlying paying non-zero dividend yield and having a general form of floating strike averaging. In a more detail,  we discuss geometric, arithmetic and weighted arithmetic averaging operator.

The pricing model is based on the assumption on the stochastic behavior of the underlying asset in time. Throughout the paper we shall assume that  the underlying asset price $S_t$ is driven by a stochastic process satisfying the following stochastic differential equation
	\begin{equation}
	\label{stoch_dif_eq}
		dS_t=(r-q)S_t\ dt+\sigma S_t\ dW_t^P,\quad 0\le t\le T.
	\end{equation}
It starts almost surely from the initial price $S_0>0$. Here the constant parameter $r>0$ denotes the risk-free interest rate whereas $q\geq0$ is a continuous dividend rate. The constant parameter $\sigma$ stands for the volatility of the underlying asset returns and $W_t^P$ is a standard Brownian motion with respect to the standard risk-neutral probability measure $P$. A solution to equation (\ref{stoch_dif_eq}) corresponds to the geometric Brownian motion
\begin{equation*}
S_t=S_0e^{(r-q-\frac12\sigma^2)t+\sigma W_t^P}, \quad 0\le t\le T.
\end{equation*}
We shall derive an integral equation determining the value of an American-style Asian option with a floating strike. If we define the optimal stopping time as $T^*$, the pay-off of the option is set by
\begin{equation*}
V_{T^*}=\Big(\rho(S_{T^*}-A_{T^*})\Big)^+,
\end{equation*}
where $V_t$ is the value of the option at time $t$, $A_t$ is a continuous average of the underlying asset value during the interval $[0,t]$ and $\rho=1$ for a call option and $\rho=-1$ for a put option. We may consider several different types of continuous averages presented in Table~\ref{averages}.
	
\begin{table}[hb]
\scriptsize
\begin{center}
\caption{\small\label{averages}Classification of averaging methods.}
\begin{tabular}{c|c|c}
arithmetic average&geometric average&weighted arithmetic average\\
\hline\hline
&&\\
$A_t=\frac1t\int_0^t{S_u}\,du$&
$\ln{A_t}=\frac1t\int_0^t{\ln{S_u}}\,du$&
$A_t=\frac1{\int_0^t{a(s)}\ ds}\int_0^t{a(t-u)S_u}\,du$
\end{tabular}
\end{center}
\end{table}
In the case of a weighted arithmetically averaged floating strike Asian option the kernel function $a(.)\geq0$ with the property $\int_0^{\infty}{a(s)}\ ds<\infty$ is usually defined as $a(s)=e^{-\lambda s}$ where $\lambda>0$ is constant.

According to \cite{ATH-PLJ-2000}, the American-style contingent claims can be priced by the conditioned expectations approach. The option price can be calculated by considering all possible stopping times in the interval $[t,T]$
\begin{equation*}
V(t,S,A)=\hbox{ess}\!\!\!\!\sup_{s\in\mathcal{T}_{[t,T]}}\ \mathbb{E}_t^P\Big[e^{-r(s-t)}\Big(\rho(S_s-A_s)\Big)^+\Big|S_t=S, A_t=A\Big],
\end{equation*}
where $\mathcal{T}_{[t,T]}$ denotes the set of all stopping times in the interval $[t,T]$ and ${\mathbb E}_t^P[X]={\mathbb E}^P[X|\mathcal{F}_t]$ is the conditioned expectation with information available at time $t$ (the information set is represented by the filtration $\mathcal{F}_t$ of the $\sigma$-algebra $\mathcal{F}$ where the Brownian motion is supported). To simplify the formula we change the probability measure by the martingale
\begin{equation*}
\eta_t=e^{-(r-q)t}\frac{S_t}{S_0}=e^{-\frac12\sigma^2t+\sigma W_t^P}
\end{equation*}
the new probability measure $\mathcal{Q}$ being defined as $d\mathcal{Q}=\eta_T\,dP$. According to Girsanov's theorem \cite{DR-MY-2005}, the process
\begin{equation*}
W_t^{\mathcal{Q}}=W_t^P-\sigma t
\end{equation*}
is a standard Brownian motion with respect to the measure $\mathcal{Q}$. The value of the underlying asset price under this measure is defined by
\begin{equation}
\label{stock_new}
S_t=S_0e^{(r-q+\frac12\sigma^2)t+\sigma W_t^{\mathcal{Q}}}.
\end{equation}		
All assets priced under this measure are $\mathcal{Q}$-martingales when discounted by the underlying price. According to this fact, we can reduce the dimension of stochastic variables. We introduce a new  variable $x_t=\frac{A_t}{S_t}$. We obtain: 
\begin{eqnarray*}
V(t,S,A)&=&\hbox{ess}\!\!\!\!\sup_{\!\!\!\!\!\!s\in\mathcal{T}_{[t,T]}}\ \mathbb{E}_t^P\Big[e^{-r(s-t)}\Big(\rho(S_s-A_s)\Big)^+
						\Big|S_t=S, A_t=A\Big]\nonumber\\
&=&\hbox{ess}\!\!\!\!\sup_{s\in\mathcal{T}_{[t,T]}}\ \mathbb{E}_t^\mathcal{Q}
\Big[\frac{\eta_t}{\eta_T}e^{-r(s-t)}\Big(\rho(S_s-A_s)\Big)^+\Big|S_t=S, A_t=A\Big]\nonumber\\
&=&\hbox{ess}\!\!\!\!\sup_{s\in\mathcal{T}_{[t,T]}}\ \mathbb{E}_t^\mathcal{Q}
\Big[e^{-q(s-t)}S_t \Big(\rho\Big(1-\frac{A_s}{S_s}\Big)\Big)^+\Big|S_t=S, A_t=A\Big]\nonumber\\
&=&\hbox{ess}\!\!\!\!\sup_{s\in\mathcal{T}_{[t,T]}}\ e^{-q(s-t)}S\ \mathbb{E}_t^\mathcal{Q}\Big[\Big(\rho(1-x_s)\Big)^+
\Big|S_t=S, A_t=A\Big].
\end{eqnarray*}
The last expression can be rewritten in terms of the new variable $x=\frac{	A}{S}$ as follows:
\begin{equation}
\label{valII}
\widetilde{V}(t,x)=e^{-qt}\frac{V(t,S,A)}{S}=e^{-q T^*_t}\ \mathbb{E}_t^\mathcal{Q}\Big[\Big(\rho(1-x_{T^*_t})\Big)^+\Big],
\end{equation}
where $T^*_t=\inf\{s\in[t,T]|x_s=x_s^*\}$ and the function $[0,T]\ni t\mapsto x^*_t\in \R $ describes the early exercise boundary.

\begin{definition}
The stopping region $\mathcal{S}$ and continuation region $\mathcal{C}$ for American-style Asian call and put options (\ref{valII}) are defined by
\begin{eqnarray*}
\mathcal{S}_{call}=\mathcal{C}_{put}=\{(t,x)| t\in [0,T], 0\leq x<x^*_t\},\ 
\mathcal{C}_{call}=\mathcal{S}_{put}=\{(t,x)| t\in [0,T], x^*_t<x<\infty\}.
\end{eqnarray*}
where  $[0,T]\in t\mapsto x^*_t\in\R$ is a continuous function determining the early exercise boundary. 
By ${\bf1}_\mathcal{S}(\cdot)$ we shall denote the indicator function of the set $\mathcal{S}$, i.e. ${\bf1}_\mathcal{S}(t,y)=1$ for $(t,y)\in  \mathcal{S}$ and ${\bf1}_\mathcal{S}(t,y)=0$ otherwise.
\end{definition}

In the following theorem we present a solution to the pricing problem with one stochastic variable $x_t$ formulated in (\ref{valII}). It is a generalization of the result by \cite{ATH-PLJ-2000} and  \cite{LW-YKK-HY-1999} for the case of a nontrivial dividend rate $q\geq0$ and a general form of the averaging of the floating strike price. 

\begin{theorem}\label{gen_val_theor}
The value $\widetilde{V}(t,x_t)$ of the American-style floating strike Asian call ($\rho=1$) or put option ($\rho=-1$) on the underlying asset $x_t$ paying continuous dividends with a rate $q\geq0$ is given by
$\widetilde{V}(t,x_t)=\widetilde{v}(t,x_t)+\widetilde{e}(t,x_t)$, where
\begin{eqnarray}
&&\widetilde{v}(t,x_t)\equiv \mathbb{E}_t^\mathcal{Q}\Big[e^{-qT}\Big(\rho(1-x_T)\Big)^+\Big],
\label{eu_price}
\\
&&\widetilde{e}(t,x_t)\equiv \mathbb{E}_t^\mathcal{Q}\Big[
\int_t^T{\rho e^{-qu}x_u {\bf1}_{\mathcal{S}}(u,x_u)\Big(\frac{dA_u}{A_u}-(r-qx_u^{-1})du\Big)}\Big],
\label{am_price}
\end{eqnarray}
with the average given by the function $A_t$ and stopping region $\mathcal{S}$. 
\end{theorem}

In the proof of Theorem~\ref{gen_val_theor} we shall  use the next lemma.
\begin{lemma}\label{lema_dif}
The auxiliary variable $x_t=\frac{A_t}{S_t}$ satisfies the following stochastic differential equation:
\begin{equation*}
dx_t=x_t\frac{dA_t}{A_t}-(r-q)x_t\ dt-\sigma x_t\ dW_t^{\mathcal{Q}}.
\end{equation*}
\end{lemma}

\medskip
\noindent P r o o f: [{\rm of Lemma~\ref{lema_dif}}]
We express the differential $dx_t=d\Big(\frac{A_t}{S_t}\Big)$ as
\[
dx_t=\frac1{S_t}dA_t-\frac{A_t}{S_t^2}dS_t+\frac{A_t}{S_t^3}(dS_t)^2
=x_t\frac{dA_t}{A_t}-(r-q)x_t\ dt-\sigma x_t\ dW_t^{\mathcal{Q}},
\]
and the proof of lemma follows.
\hfill$\square$\medskip

Notice that, when comparing to the original expression due to \cite{ATH-PLJ-2000} with a zero dividend rate $q=0$ the only difference is that the parameter $r$ is replaced by the term $r-q$.

The value of $\frac{dA_t}{A_t}$ depends on the method of averaging of the underlying asset used in valuation. The expressions for continuous averages are presented in Table~\ref{AA_hodnoty}. In this table we present the value of the weighted arithmetic average restricted to the kernel $a(s)=e^{-\lambda s}$.	
	
\begin{table}[hb]
\begin{center}
\caption{\small\label{AA_hodnoty}The value of the differential $\frac{dA_t}{A_t}$.}
\scriptsize
\begin{tabular}{c|c|c}
arithmetic average&geometric average&weighted arithmetic average\\
\hline\hline
			&&\\
			$\frac{dA^a_t}{A^a_t}=\frac1t\Big(\frac1{x^a_t}-1\Big)\ dt$&
				$\frac{dA^g_t}{A^g_t}=-\frac1t\ln{x^g_t}\ dt$&
					$\frac{dA^{wa}_t}{A^{wa}_t}=\frac{\lambda}{1-e^{-\lambda t}}\Big(\frac1{x^{wa}_t}-1\Big)\ dt$\\
			&&
		\end{tabular}
		\end{center}
	\end{table}

\smallskip
\noindent P r o o f: [{\rm of Theorem~\ref{gen_val_theor}}]
We follow the proof of the original result by \cite{ATH-PLJ-2000} and we include necessary modifications related to the form of averaging and the fact that $q\geq0$.

First, we suppose that $(t,x)\in\mathcal{C}$. The option is held and so we can apply It\=o's lemma to calculate the differential
\begin{eqnarray*}
			d\widetilde{V}&=&\frac{\partial\widetilde{V}}{\partial x}dx+\frac12\frac{\partial^2\widetilde{V}}{\partial x^2}(dx)^2
						+\frac{\partial\widetilde{V}}{\partial t}dt\\
					&=&x\frac{\partial\widetilde{V}}{\partial x}\frac{dA}{A}
							+\Big[-(r-q)x\frac{\partial\widetilde{V}}{\partial x}
							+\frac12\sigma^2 x^2\frac{\partial^2\widetilde{V}}{\partial x^2}
							+\frac{\partial\widetilde{V}}{\partial t}\Big]dt
							-\sigma x\frac{\partial\widetilde{V}}{\partial x}dW^{\mathcal{Q}}\\
					&=&-\sigma x\frac{\partial\widetilde{V}}{\partial x}dW^{\mathcal{Q}},
\end{eqnarray*}
where the last equality holds true, because $\widetilde{V}$ is $\mathcal{Q}$-martingale.

Now we suppose that $(t,x)\in\mathcal{S}$. The value of the option is defined by
		\[
			\widetilde{V}(t,x_t)=\rho e^{-qt}(1-x_t).
		\]
		Hence the differential $d\widetilde{V}=-\rho q e^{-qt}(1-x)dt-\rho e^{-qt}dx$ has the form
		\[
			d\widetilde{V}=-\rho e^{-qt}x\frac{dA}{A}+\rho e^{-qt}(rx-q)dt+\rho e^{-qt}\sigma x dW^{\mathcal{Q}}.
		\]
		For both regions we have the following equation
		\begin{equation}
			\label{val_pr_I}
			d\widetilde{V}(t,x_t)=-\rho e^{-qt}{\bf1}_{\mathcal{S}}(t,x_t)\Big(x_t\frac{dA_t}{A_t}-(rx_t-q)dt\Big)+dM_t^{\mathcal{Q}},
		\end{equation}
		where $M_t^{\mathcal{Q}}$ is a $\mathcal{Q}$-martingale. Integrating (\ref{val_pr_I}) from $t$ to $T$ and taking expectation we have
		\begin{eqnarray*}
			\mathbb{E}_t^\mathcal{Q}\Big[\widetilde{V}(T,x_T)\Big]-\widetilde{V}(t,x_t)&=&
					-\mathbb{E}_t^\mathcal{Q}\Big[\int_t^T{\rho e^{-q u}x_u {\bf1}_{\mathcal{S}}(u,x_u)\Big(\frac{dA_u}{A_u}-(r-\frac{q}{x_u})du \Big)}\Big]\\
					&&+\underbrace{\mathbb{E}_t^\mathcal{Q}\Big[\int_t^T{dM_u^{\mathcal{Q}}}\Big]}_{=0},
		\end{eqnarray*}
		\begin{eqnarray*}
			\widetilde{V}(t,x_t)&=&\underbrace{\mathbb{E}_t^\mathcal{Q}\Big[e^{-q T}\Big(\rho(1-x_T)\Big)^+\Big]}_{=\widetilde{v}(t,x_t)}
					+\underbrace{\mathbb{E}_t^\mathcal{Q}\Big[\int_t^T{\rho e^{-q u}x_u {\bf1}_{\mathcal{S}}(u,x_u)\Big(\frac{dA_u}{A_u}
							-(r-\frac{q}{x_u}) du\Big)}\Big]}_{=\widetilde{e}(t,x_t)},
		\end{eqnarray*}		
which completes the proof of Theorem~\ref{gen_val_theor}.
	\hfill$\square$\medskip

\smallskip
It is worthwhile noting that the above expression for the value $\widetilde{V}(t,x_t)=\widetilde{V}_{am}(t,x_t)$ of an American-style Asian option can be restated as follows: 
\[
\widetilde{V}_{am}(t,x) = \widetilde{V}_{eu}(t,x) + \mathbb{E}_t^\mathcal{Q}\Big[\int_t^T {\bf1}_{\mathcal{S}}(u,x_u) f_b(u,x_u) du \Big]
\]
where $\widetilde{V}_{eu}(t,x) = \mathbb{E}_t^\mathcal{Q}\Big[e^{-q T}\Big(\rho(1-x_T)\Big)^+\Big]$ stands for the price of the European-style Asian option and the term ${\bf1}_{\mathcal{S}}(u,x_u) f_b(u,x_u), u\in [0,T],$ represents a surplus bonus for the difference between American and European-style of the Asian option. 

\section{Early exercise behavior close to expiry}
	
	\subsection{Limit of the early exercise boundary at expiry}
		
		In this section we determine the position of the early exercise boundary $x^*_T$ at expiry $T$. The result is stated for a wide class of integral equations for pricing American-style options.

		\begin{theorem}
			\label{st_p}
			Consider an American-style (call or put) option $V_{am}$ on the underlying $y$ with the stopping and continuation regions defined by the sets $\mathcal{S}\not=\emptyset$, and $\mathcal{C}\not=\emptyset$, resp. Let $y^*_t=\partial\mathcal{S}(t,.)\equiv \partial\mathcal{C}(t,.)$  for $t\in[0,T]$ be the early exercise boundary function. Suppose that the value of $V_{am}$ is given by the equation 
			\begin{equation}
			\label{americ}
				V_{am}(t,y_t)=V_{eu}(t,y_t)					+\mathbb{E}_t\Big[\int_t^T{{\bf1}_{\mathcal{S}}(u,y_u)f_{\textrm{b}}(u,y_u)}\ du\Big],
			\end{equation}
			where $V_{eu}$ denotes the price of the corresponding European-style option and $f_{\textrm{b}}(t,y)$ is a continuous function representing the early exercise bonus such that the equation $f_{\textrm{b}}(T,y)=0$ has a unique root $y^*\ge0$. Furthermore we suppose that  
\[
V_{am}(t,y)\ge V_{am}(T,y) = V_{eu}(T,y)\ge0, V_{am}(t,y)\ge V_{eu}(t,y) \quad \hbox{for any}\ t\in [0,T], y\ge 0,
\]
and the function $[0,\infty)\ni y\mapsto \frac{\partial V_{eu}}{\partial t} (T,y)\in \R$ is continuous except of the set $ATM=\partial\{ y\ge0, V_{eu}(T,y) >0\}$. Then only one of the following cases can occur:
\begin{enumerate}
\item $y^*_T\in ITM=\{ y\ge0, V_{eu}(T,y) >0\}$. In this case $f_{\textrm{b}}(T,y^*_T)=0$.
\item $y^*_T\in ATM=\partial\{ y\ge0, V_{eu}(T,y) >0\}$. In this case  $f_{\textrm{b}}(T,y^*_T)\ge 0$.

\end{enumerate}
	
		\end{theorem}

\begin{remark}
The abbreviation {\it ITM} stands for the so-called in-the-money set whereas {\it ATM} (the boundary of {\it ITM}) denotes  the so-called at-the-money set. We denote by  {\it OTM} the out-the-money set - the complement of sets {\it ITM} and {\it ATM}, i.e. $OTM = (ITM \cup ATM)^c$. 
\end{remark}

\medskip

\noindent P r o o f: [{\rm of Theorem~\ref{st_p}}]
We have
\[
\frac{1}{T-t}\mathbb{E}_t\Big[\int_t^T{{\bf1}_{\mathcal{S}}(u,y_u)f_{\textrm{b}}(u,y_u)}\ du\Big]
= \frac{1}{T-t}\left( V_{am}(t,y_t)-V_{eu}(t,y_t) \right) \ge 0,
\]
for any $t\in [0,T)$. In the limit $t\to T$, we can omit the conditioned expected value operator $\mathbb{E}_t$ and we obtain
$
{\bf1}_{\mathcal{S}}(T,y_T)f_{\textrm{b}}(T,y_T)\geq0. 
$
Since $(T,y^*_T) \in \partial S$ and the function $f_{\textrm{b}}$ is continuous we obtain 
$f_{\textrm{b}}(T,y^*_T)\ge 0$.

\smallskip
Part 1). 
Suppose that $y^*_T\in ITM$. We shall prove $f_{\textrm{b}}(T,y^*_T) = 0$. 
Notice that in the stopping region $\mathcal{S}$ we have the identity $V_{am}(t,y) = V_{am}(T,y)$ for any $(t,y)\in \mathcal{S}$ and, consequently, $\frac{\partial V_{am}}{\partial t} (t,y)=0$. Take any $(T,y_T)\in \mathcal{S}$. Similarly as in above, in the limit $t\to T$ we can omit the conditioned expected value operator $\mathbb{E}_t$ to obtain 
\[
0=\frac{\partial V_{am}}{\partial t} (T,y_T)= \frac{\partial V_{eu}}{\partial t} (T,y_T) - f_{\textrm{b}}(T,y_T). 
\]
Since the function $y\mapsto \frac{\partial V_{eu}}{\partial t}(T,y)$ is assumed to be continuous in {\it ITM} and $y^*_T\in ITM$, we have $\frac{\partial V_{eu}}{\partial t} (T,y^*_T) = f_{\textrm{b}}(T,y^*_T)$. 

On the other hand, take any $(T,y_T)\in \mathcal{C}$. Since $V_{am}(t,y)\ge V_{am}(T,y)$ for any $0\le t<T$ we obtain 
$\frac{\partial V_{am}}{\partial t} (T,y_T)\le 0$. Therefore, in the limit $t\to T$, we have
\[
0\ge \frac{\partial V_{am}}{\partial t} (T,y_T)= \frac{\partial V_{eu}}{\partial t} (T,y_T) - {\bf 1}_{\mathcal{S}}(T,y_T) f_{\textrm{b}}(T,y_T) = \frac{\partial V_{eu}}{\partial t} (T,y_T)
\]
because $(T,y_T)\in \mathcal{C}$. Due to continuity of $\frac{\partial V_{eu}}{\partial t}(T,y)$ for $y\in ITM$ we finally obtain $0\ge\frac{\partial V_{eu}}{\partial t} (T,y^*_T)= f_{\textrm{b}}(T,y^*_T)$. Hence $f_{\textrm{b}}(T,y^*_T)=0$ and the proof of Part 1) follows. 

\smallskip
Part 2). It suffices to prove that the case $y^*_T\in OTM$ cannot occur. Suppose to the contrary $y^*_T\in OTM$. Since the set {\it OTM} is open we can argue similarly as in the first part of the proof to obtain $f_{\textrm{b}}(T,y^*_T)=0$. Since we have assumed uniqueness of the root of the equation $f_{\textrm{b}}(T,y)=0$ then, in some neighborhood of $y^*_T$ there exists $\widehat{y}_T\in OTM $ such that $f_{\textrm{b}}(T,\widehat{y}_T)>0$ and $(T,\widehat{y}_T)\in \mathcal{S}$.  We have
\[
0=\frac{\partial V_{am}}{\partial t}(T,\widehat{y}_T)=\frac{\partial V_{eu}}{\partial t}(T,\widehat{y}_T)-f_{\textrm{b}}(T,\widehat{y}_T).
\]
Thus $\frac{\partial V_{eu}}{\partial t}(T,\widehat{y}_T)>0$. For any $y\in OTM$ we have $V_{eu}(T,y)=0$. Hence $V_{eu}(T,\widehat{y}_T)=0$ and $V_{eu}(t,\widehat{y}_T)\geq0$ for all $0\le t<T$, a contradiction. Thus $y^*_T\not\in OTM$ and the proof of theorem follows.
		\hfill$\square$\medskip

As a consequence of Theorem~\ref{st_p} we obtain the starting position of the early exercise boundary for American-style of Asian options with various types of the strike price averaging method. 

\begin{corollary}
The value of the limit of early exercise boundary at expiry for the floating strike Asian option is summarized in the Table~\ref{start_point}. In the case of a geometric averaging, it follows from (\ref{am_price}) that $\widetilde{f}_{\textrm{b}}(T,x_T)=e^{-qT} \big(-\frac{x_T}T\ln{x_T}-rx_T+q\big)$ such that $x^*_T=\widetilde{x}_T\in ITM$ is a solution  of the transcendent equation
\begin{equation}
	\label{transc_eq}
	\ln{\widetilde{x}_T}=\frac{qT}{\widetilde{x}_T}-rT.
\end{equation}
\end{corollary}
The formula for the limit of the early exercise boundary at the expiry (\ref{transc_eq}) for geometric averaging is the same as presented by \cite{LW-YKK-HY-1999} and \cite[p. 69]{JD-2006}. Notice that the same values of the limit of early exercise boundary at expiry for the continuous arithmetic average type of an Asian option is derived also in \cite{MD-YKK-2006}.

\begin{table}[h]
\scriptsize
\begin{center}
\caption{\small\label{start_point}The limit of the early exercise boundary position $x_T^*$ at expiry $t=T$ ($\widetilde{x}_T$ solves (\ref{transc_eq})).}
\begin{tabular}{c||c|c|c}
$x^*_T$&arithmetic average&geometric average&weighted arithmetic average\\
				\hline\hline
				put&
					$\max{\Big(\frac{q+\frac1T}{r+\frac1T},1\Big)}$&
						$\max{\Big(\widetilde{x}_T,1\Big)}$&
							$\max{\Big(\frac{q (1-e^{-\lambda t})+\lambda}{r (1-e^{-\lambda t})+\lambda},1\Big)}$\\
				\hline
				call&
					$\min{\Big(\frac{q+\frac1T}{r+\frac1T},1\Big)}$&
						$\min{\Big(\widetilde{x}_T,1\Big)}$&
							$\min{\Big(\frac{q (1-e^{-\lambda t})+\lambda}{r (1-e^{-\lambda t})+\lambda},1\Big)}$
			\end{tabular}
			\end{center}
		\end{table}

		\begin{figure}
			\begin{center}
				\includegraphics[width=0.3\textwidth]{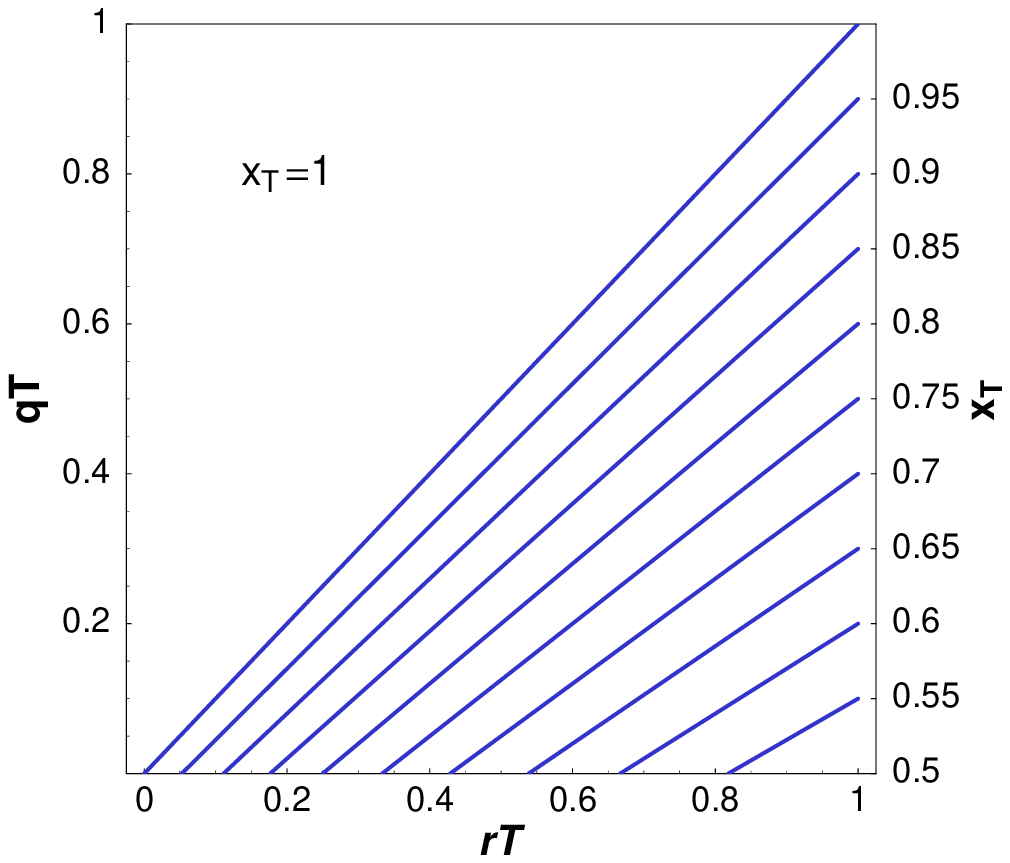}
				\includegraphics[width=0.3\textwidth]{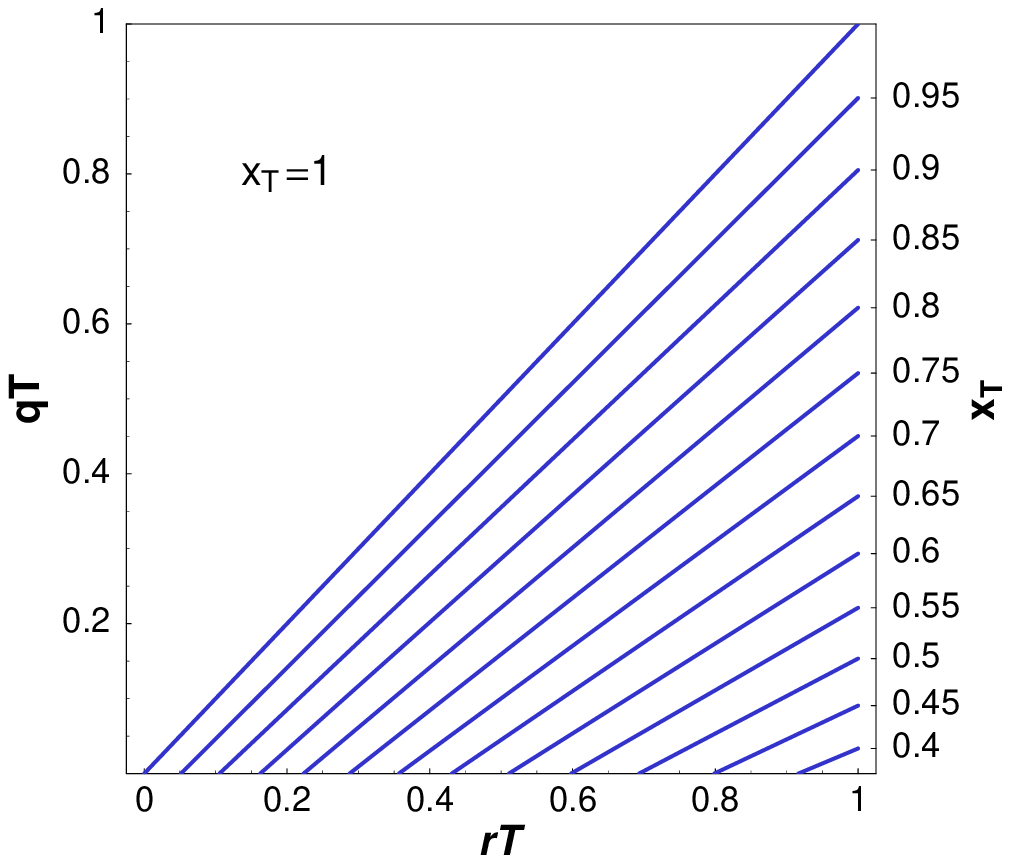}
			\end{center}
\vglue-0.5truecm
			\caption{\small  Isolines of the the limit $x^*_T$ of the early exercise boundary at expiry $T$ of call option for the continuous arithmetic (left) and the continuous geometric average (right).}
			\label{obr-ktabulke}
		\end{figure}

	\subsection{Integral equation for pricing Asian options}

		In this section we calculate the approximate formula for the American-style Asian option with various floating strike averages. The next lemma will be useful in calculations to follow.

		\begin{lemma}\cite{ATH-PLJ-2000}
			\label{strhod}
			Let $\omega=\ln{\Omega}\sim \mathcal{N}(\alpha,\beta^2)$
			and define $\gamma\equiv \frac{\alpha+\beta^2-\ln{K}}{\beta}$, where $K>0$. We have
			\begin{eqnarray*}
				&&\mathbb{E}\big[1_{\{\rho\Omega\geq\rho K\}}\big]={\bf\Phi}(\rho(\gamma-\beta)),\,
				\mathbb{E}\big[1_{\{\rho\Omega\geq\rho K\}}\Omega\ln{\Omega}\big]=
							e^{\alpha+\frac{\beta^2}{2}}\Big((\alpha+\beta^2){\bf\Phi}(\rho\gamma)+\rho\beta{\it\Phi}(\gamma)\Big),\\
				&&\mathbb{E}\big[1_{\{\rho\Omega\geq\rho K\}}\Omega\big]=e^{\alpha+\frac{\beta^2}{2}}{\bf\Phi}(\rho\gamma),\,
				\mathbb{E}\big[(\rho(\Omega-K))^+\big]=\rho\Big(e^{\alpha+\frac{\beta^2}{2}}{\bf\Phi}(\rho\gamma)-K{\bf\Phi}(\rho(\gamma-\beta))\Big),
			\end{eqnarray*}
			where $\rho\in\{-1,1\}$ and ${\bf\Phi}(\cdot)$ and ${\it\Phi}(\cdot)$ are standard normal cumulative distribution and density functions, resp.
		\end{lemma}

		\subsubsection{Geometric average}

In this section we recall the integral equation for pricing American-style of Asian geometrically averaged floating strike options. It was derived for the case $q=0$ by \cite{ATH-PLJ-2000} and for the general case $q\ge 0$ by \cite{LW-YKK-HY-1999}. 

\begin{lemma}\cite{LW-YKK-HY-1999}
				\label{geom_momenty}
				In the case of geometric averaging, the variable $x^g_t=\frac{A_t^g}{S_t}$ has log-normal (conditioned) distribution
				$\ln{x_u^g}|\mathcal{F}_t\sim \mathcal{N}(\alpha_{t,u},\beta^2_{t,u})$, where $u\geq t$ and parameters
				$\alpha_{t,u}=\alpha(t,u,x_t)$ and $\beta_{t,u}=\beta(t,u)$ are defined by
				$$
					\alpha^g_{t,u}=\frac{t}{u}\ln{x_t^g}-\frac{u^2-t^2}{2u}(r-q+\frac{\sigma^2}{2}),\quad
					\beta^g_{t,u}=\frac{\sigma}{u\sqrt{3}}\sqrt{u^3-t^3}.
				$$
			\end{lemma}

Now one can apply  Lemma~\ref{strhod} in order to calculate the formula for option with the geometric averaging. Recall that for the floating strike Asian call or put option, the stopping region $\mathcal{S}= \{ (t,x), x\ge 0, \rho x^*_t > \rho x\}$, where $x^*_t$ is the exercise boundary and $\rho=1$ for the case of a call option whereas $\rho=-1$ for a put option. If we insert the expression $\frac{dA^g_t}{A^g_t}$ 
for the geometric average (see Table~\ref{averages}) into (\ref{eu_price}) and (\ref{am_price}) we obtain the formula for the European-style option
			\begin{eqnarray*}
				\widetilde{v}^g(t,x_t)&=&\mathbb{E}_t^{\mathcal{Q}}\Big[e^{-q T}\Big(\rho(1-x^g_T)\Big)^+\Big]
					=e^{-q T}\mathbb{E}_t^{\mathcal{Q}}\Big[\Big(\rho(1-x^g_T)\Big)^+\Big]\nonumber\\
				&=&\rho e^{-q T}\Bigg({\bf\Phi}\Big(-\rho\Big(\frac{\alpha^g_{t,T}}{\beta^g_{t,T}}\Big)\Big)
						-e^{\alpha^g_{t,T}+\frac{(\beta^g_{t,T})^2}2}
						{\bf\Phi}\Big(-\rho\Big(\frac{\alpha^g_{t,T}}{\beta^g_{t,T}}+\beta^g_{t,T}\Big)\Big)\Bigg).
			\end{eqnarray*}
and the value of the American early exercise bonus premium 
			\begin{eqnarray*}
				\widetilde{e}^g(t,x_t)&=&\mathbb{E}_t^\mathcal{Q}\Big[
						\int_t^T{\rho e^{-q u}x^g_u
					{\bf1}_{\mathcal{S}}(u,x^g_u)\Big(\frac{dA^g_u}{A^g_u}-(r-q(x^g_u)^{-1})du\Big)}\Big]\nonumber\\
			&=&\int_t^T{\rho e^{-q u}\mathbb{E}_t^\mathcal{Q}\Big[ {\bf1}_{\{\rho x^*_u > \rho x^g_u\}}
					\Big(-\frac1u x^g_u\ln{x^g_u}-rx^g_u+q\Big)\Big]}du\\
					&=&\int_t^T{\rho e^{-q u}\Bigg(q{\bf\Phi}(\rho(\beta^g_{t,u}-\gamma^g_{t,u}))}\nonumber\\
				&&{+e^{\alpha^g_{t,u}+\frac{(\beta^g_{t,u})^2}2}
					\Bigg(\rho\frac{\beta^g_{t,u}}{u}{\bf\Phi}'(\gamma^g_{t,u})-\Big(r+\frac{\alpha^g_{t,u}
					+(\beta^g_{t,u})^2}{u}\Big){\bf\Phi}(-\rho\gamma^g_{t,u})\Bigg)\Bigg)}du\nonumber,
			\end{eqnarray*}
			where ${\bf\Phi}(\cdot)$ is the standard normal cumulative distribution function and
			\begin{equation*}
				\gamma^g_{t,u}=\frac{\alpha^g_{t,u}-\ln{x^*_u}}{\beta^g_{t,u}}+\beta^g_{t,u}.
			\end{equation*}

			Returning to the original variables we obtain the formula of American-style floating strike Asian option with geometrically
			averaged floating strike:
			\begin{equation*}
				V^g(t,S,A)=S e^{qt}\widetilde{V}^g(t,A/S)=S e^{qt}(\widetilde{v}^g(t,A/S)+\widetilde{e}^g(t,A/S)).
			\end{equation*}
			If we formally set value of the continuous dividend rate to zero, i.e. $q=0$, the result is identical to the expression obtained in 
			the paper \cite{ATH-PLJ-2000}).

		\subsubsection{Approximation for the arithmetic average}
			\label{HJ_bias}

			Unfortunately, in the case of an arithmetically averaged floating strike Asian option the probabilistic distribution function of the arithmetic average cannot be expressed in an explicit way. Following \cite{ATH-PLJ-2000} we approximate
			the probabilistic distribution of the variable $x_t^a=\frac{A_t^a}{S_t}$ for the continuous arithmetic average $A_t^a$ by the log-normal conditioned distribution, i.e. 
			$\ln{x_u^a}|\mathcal{F}_t\sim \mathcal{N}(\alpha^a_{t,u},(\beta^a_{t,u})^2)$ at time $t$, where
			\begin{equation}
				\label{arit_ab}
				\alpha^a_{t,u}=2\ln{\mathbb{E}_t^{\mathcal{Q}}\big[x^a_u\big]}-\frac12\ln{\mathbb{E}_t^{\mathcal{Q}}\big[(x^a_u)^2\big]},\quad
				\beta^a_{t,u}=\sqrt{\ln{\mathbb{E}_t^{\mathcal{Q}}\big[(x^a_u)^2\big]}-2\ln{\mathbb{E}_t^{\mathcal{Q}}\big[x^a_u\big]}}.
			\end{equation}

			\begin{lemma}
				\label{arit_momenty}
				Consider the variable $x_u=\frac{A_u}{S_u}$, where $A_u$ and $S_u$ are defined as the arithmetic average
(see Table~\ref{averages}) and as in (\ref{stock_new}), resp. First two conditioned moments 
$\mathbb{E}_t^{\mathcal{Q}}\big[x^a_u\big]$ and $\mathbb{E}_t^{\mathcal{Q}}\big[(x^a_u)^2\big]$ of $x_u$
entering the expressions for the functions $\alpha^a_{t,u}=\alpha^a(t,u,x^a_t)$ and $\beta^a_{t,u}=\beta^a(t,u,x^a_t)$ can be calculated, for $t\leq u$, as follows:
				\begin{eqnarray}
					\mathbb{E}_t^{\mathcal{Q}}\big[x^a_u\big]&=&x^a_t\frac{t}{u}e^{-(r-q)(u-t)}+\frac1{(r-q)u}\big(1-e^{-(r-q)(u-t)}\big),\\
					\label{druhy_moment}
						\mathbb{E}_t^{\mathcal{Q}}\big[(x^a_u)^2\big]&=&(x^a_t)^2\frac{t^2}{u^2}e^{-2(r-q-\frac{\sigma^2}2)(u-t)}
							+x^a_t\frac{2 t e^{-(r-q)(u-t)}}{u^2(r-q)}\big(1-e^{-(r-q)(u-t)}\big)\\
							&&+\frac{(r-q-\sigma^2)-2(r-q-\frac{\sigma^2}2)e^{-(r-q)(u-t)}+(r-q)e^{-2(r-q-\frac{\sigma^2}2)(u-t)}}
								{u^2(r-q)(r-q-\frac{\sigma^2}2)(r-q-\sigma^2)}.\nonumber
				\end{eqnarray}
			\end{lemma}
			\medskip
			\noindent P r o o f: 
				Following the lines of the derivation of $\mathbb{E}_t^{\mathcal{Q}}[x^a_u]$ from \cite{ATH-PLJ-2000} adopted for a general dividend yield $q\ge 0$ we obtain the first moment
				\[
					\mathbb{E}_t^{\mathcal{Q}}\Big[x^a_u\Big]=x^a_t\frac{t}{u}e^{-(r-q)(u-t)}+\frac{1-e^{-(r-q)(u-t)}}{(r-q)u}.
				\]
				Although we follow the proof by \cite{ATH-PLJ-2000}, we have to make a slight correction in the derivation of the second
				conditioned moment.
				Using the definition of $S_t$, we have, for all $v\in [t,u]$,
				\[
					\mathbb{E}_t^{\mathcal{Q}}\Big[\frac{S_v}{S_u}\Big]=
					\mathbb{E}_t^{\mathcal{Q}}\Big[e^{(r-q+\frac{\sigma^2}2)(v-u)+\sigma(W_v^{\mathcal{Q}}-W_u^{\mathcal{Q}})}\Big]=
					e^{(r-q+\frac{\sigma^2}2)(v-u)+\frac{\sigma^2}2|v-u|}=e^{(r-q)(v-u)}.
				\]
				We need to simplify the expression for the the second conditioned moment
				\begin{eqnarray*}
					&&\mathbb{E}_t^{\mathcal{Q}}\Big[(x^a_u)^2\Big]
					=\mathbb{E}_t^{\mathcal{Q}}\Big[\Big(\frac1u\int_0^u{\frac{S_v}{S_u}}dv\Big)^2\Big]\\
					&=&(x^a_t)^2\frac{t^2}{u^2}\mathbb{E}_t^{\mathcal{Q}}\Big[\frac{S_t}{S_u}\frac{S_t}{S_u}\Big]
						+2x_t\frac{t}{u^2}
						\mathbb{E}_t^{\mathcal{Q}}\Big[\frac{S_t}{S_u}\Big]\int_t^u{\mathbb{E}_t^{\mathcal{Q}}\Big[\frac{S_v}{S_u}\Big]}dv
						+\frac1{u^2}\int_t^u\int_t^u{\mathbb{E}_t^{\mathcal{Q}}\Big[\frac{S_z}{S_u}\frac{S_v}{S_u}\Big]}dv\ dz.
				\end{eqnarray*}
				Assuming that $u\geq z,v$ and let $m=\min{\{z,v\}}$ and $M=\max{\{z,v\}}$, we have
				\begin{eqnarray*}
					e^{-(r-q+\frac{\sigma^2}2)(z+v-2u)}\frac{S_z}{S_u}\frac{S_v}{S_u}&=&e^{\sigma(W_z^{\mathcal{Q}}+W_v^{\mathcal{Q}}-2W_u^{\mathcal{Q}})}
						=e^{-\sigma(2(W_u^{\mathcal{Q}}-W_{M}^{\mathcal{Q}})
						+(W_{M}^{\mathcal{Q}}-W_{m}^{\mathcal{Q}}))},\\
					\mathbb{E}_t^{\mathcal{Q}}\Big[\frac{S_z}{S_u}\frac{S_v}{S_u}\Big]
						&=&e^{(r-q)(z+v-2u)+\sigma^2(u-M)}.
				\end{eqnarray*}
				We have calculated all expressions we need to evaluate the second conditioned moment. If we put all together
				and perform necessary calculation we obtain (\ref{druhy_moment}) and the proof of lemma follows.
			\hfill$\square$\medskip

\begin{remark}
If we formally set the value of the continuous dividend rate $q=0$ in Lemma~\ref{arit_momenty} we obtain almost identical expression to that of \cite{ATH-PLJ-2000} except of the second moment $\mathbb{E}_t^{\mathcal{Q}}\Big[(x^a_u)^2\Big]$ entering (\ref{arit_ab}). The expression 
			\begin{eqnarray*}
				\mathbb{E}_t^{\mathcal{Q}}\big[(x^a_u)^2\big]_{HJ}&=&(x^a_t)^2\frac{t^2}{u^2}e^{-2(r-\frac{\sigma^2}2)(u-t)}
					+x^a_t\frac{2 t e^{-r(u-t)}}{u^2(r-\sigma^2)}\big(1-e^{-(r-\sigma^2)(u-t)}\big)\\
					&&+\frac{(r-\sigma^2)-2(r-\frac{\sigma^2}2)e^{-r(u-t)}+re^{-2(r-\frac{\sigma^2}2)(u-t)}}
						{u^2 r(r-\frac{\sigma^2}2)(r-\sigma^2)}.
			\end{eqnarray*}
by  \cite{ATH-PLJ-2000} differs from  our (\ref{druhy_moment}) in the second summand where
the term $r-\sigma^2$ is replaced by $r$ in both in the denominator and exponent. 
\end{remark}

\smallskip

Now we can return to the problem of valuation of an Asian option. First we replace the general form of the average in (\ref{am_price}) by the
expression for the arithmetic average defined as arithmetic average in Table~\ref{averages}. The stopping region $\mathcal{S}$ is the same as for the case of geometric averaging.

Using Lemma~\ref{strhod} we calculate the value of both (\ref{eu_price}) and (\ref{am_price}). 
The European part of the
			option has value
			\begin{eqnarray*}
				\widetilde{v}^a(t,x)&=&\mathbb{E}_t^{\mathcal{Q}}\Big[e^{-q T}\Big(\rho(1-x^a_T)\Big)^+\Big]\nonumber\\
				&=&\rho e^{-q T}\Bigg({\bf\Phi}\Big(-\rho\Big(\frac{\alpha^a_{t,T}}{\beta^a_{t,T}}\Big)\Big)
						-e^{\alpha^a_{t,T}+\frac{(\beta^a_{t,T})^2}2}
						{\bf\Phi}\Big(-\rho\Big(\frac{\alpha^a_{t,T}}{\beta^a_{t,T}}+\beta^a_{t,T}\Big)\Big)\Bigg)
			\end{eqnarray*}
			and the American early exercise bonus premium
			\begin{eqnarray*}
				\widetilde{e}^a(t,x)&=&\mathbb{E}_t^\mathcal{Q}\Big[
						\int_t^T{\rho e^{-q u}x^a_u {\bf1}_{\mathcal{S}}(u,x^a_u)
							\Big(\frac{dA^a_u}{A^a_u}-(r-q(x^a_u)^{-1})du\Big)}\Big]\nonumber\\
				&=&\int_t^T{\rho e^{-q u}\mathbb{E}_t^\mathcal{Q}\Big[1_{\{\rho x^*_u>\rho x^a_u\}}
					\Big(\frac1u(1-x^a_u)-rx^a_u+q\Big)\Big]}du\\
					&=&\int_t^T{\rho e^{-q u}\Bigg(\Big(q+\frac1u\Big){\bf\Phi}(\rho(\beta^a_{t,u}-\gamma^a_{t,u}))
							-\Big(r+\frac1u\Big)e^{\alpha^a_{t,u}+\frac{(\beta^a_{t,u})^2}2}{\bf\Phi}(-\rho\gamma^a_{t,u})\Bigg)}du\nonumber,
			\end{eqnarray*}
			where ${\bf\Phi}(\cdot)$ is the standard normal cumulative distribution function and
			\begin{equation*}
				\gamma^a_{t,u}=\frac{\alpha^a_{t,u}-\ln{x^*_u}}{\beta^a_{t,u}}+\beta^a_{t,u}.
			\end{equation*}

Returning to the original variables we have the approximate value of American-style Asian option with a continuous arithmetic averaging
			\begin{equation*}
				V^a(t,S,A)=S e^{qt}\widetilde{V}^a(t,x)=S e^{qt}(\widetilde{v}^a(t,x)+\widetilde{e}^a(t,x)).
			\end{equation*}

	\subsection{Expansion of the early exercise boundary close to expiry}
		
Throughout this section we shall assume the structural assumption on the interest and dividend rates:
\begin{equation}
r>q\ge 0.
\label{rq-ner}
\end{equation}
We shall calculate an approximation of the call option early exercise boundary function for a call option. The approximation is obtained by the first order Taylor series expansion in the $\sqrt{\tau}$ variable, where $\tau=T-t$ is the time to expiry. To approximate the early exercise boundary function by the Taylor expansion, we need to calculate the first derivative of $x^*_t$ at
expiry $T$ with respect to $\sqrt{T-t}$ variable. Following \cite{RAK-JBK-1998,NJD-SDH-IR-PW-1993,DS-2001} we propose an approximation of the early exercise boundary $x^*_t$ in the form 
		\begin{equation*}
			x^*_t=x^*_T(1+C\sqrt{T-t})+O(T-t) \quad\hbox{as}\ t\to T,
		\end{equation*}
where $C\in\R$ is a constant. To calculate $C$, we use the condition of smoothness of the early exercise boundary of the call option across the early exercise boundary - smooth pasting principle (c.f. \cite{YKK-1998,MD-YKK-2006}). Since $\widetilde{V}(T,x) = e^{-qT} (1-x)^+$ we have 

\begin{equation}
			\label{rov_hran}
			-1=e^{qt}\frac{\partial \widetilde{V}}{\partial x}(t,x_t^*)=e^{qt}\frac{\partial\widetilde{v}_t}{\partial x}(t,x_t^*)+e^{qt}\frac{\partial\widetilde{e}_t}{\partial x}(t,x_t^*)
				=\widehat{v}_{x}(t,x_t^*)+\int_t^T{\widehat{e}_{Ix}(t,x_t^*,u,x_u^*)}\ du.
		\end{equation}
In further derivation we use following limits (according to the Lemma~\ref{geom_momenty},
		Lemma~\ref{arit_momenty} and Table~\ref{start_point}). We recall that $\alpha_{t,u}=\alpha(t,u,x)$,
		$\beta^a_{t,u}=\beta^a(t,u,x^a)$ and $\beta^g_{t,u}=\beta^g(t,u)$.
		\begin{eqnarray}
		\label{limity}				
			\lim_{\tau\rightarrow0}\alpha_{T-\tau,T}=\lim_{\tau\rightarrow0}\alpha_{T-\tau,T-\tau(1-\theta)}&=&\alpha_{T,T}=\ln{x_T^*}<0,\nonumber\\
			\lim_{\tau\rightarrow0}\beta_{T-\tau,T}=\lim_{\tau\rightarrow0}\beta_{T-\tau,T-\tau(1-\theta)}&=&\beta_{T,T}=0^+,\nonumber \\
			\lim_{\tau\rightarrow0}{\bf\Phi}\Big(-\frac{\alpha_{T-\tau,T}}{\beta_{T-\tau,T}}\Big)&=&{\bf\Phi}\Big(-\frac{\ln{x_T^*}}{0^+}\Big)=1,\nonumber\\
			\lim_{\tau\rightarrow0}\frac{{\bf\Phi}'\Big(-\frac{\alpha_{T-\tau,T}}{\beta_{T-\tau,T}}\Big)}{(\beta_{T-\tau,T})^n}&=&0,\\
			\lim_{\tau\rightarrow0}\frac{\ln{x_T^*(1+h\sigma\sqrt{\tau(1-\theta)})}-\alpha_{T-\tau,T-\tau(1-\theta)}}{\beta_{T-\tau,T-\tau(1-\theta)}}&=&-h\frac{1-\sqrt{1-\theta}}{\sqrt{\theta}},\nonumber\\
			\lim_{\tau\rightarrow0}\partial_x\alpha_{T-\tau,T}=\lim_{\tau\rightarrow0}\partial_x\alpha_{T-\tau,T-\tau(1-\theta)}&=&\frac1{x_T^*},\nonumber \\
			\lim_{\tau\rightarrow0}\beta_{T-\tau,T}\,\partial_{\tau}\beta_{T-\tau,T}&=&\frac{\sigma^2}2.\nonumber
		\end{eqnarray}
Since we have assumed $r>q\ge0$ we have $0<x_T^*<1$ (see Table~\ref{start_point}).
Notice that both $\alpha$ and $\beta$ have polynomial order in $\tau$ and the derivative of the normal
		cumulative distribution function (i.e. the probability density function) has exponential order in $\tau$ variable.
		In both derivations we have used several properties of the derivative of normal cumulative distribution function ${\bf\Phi}(x)$, e.g. 
		${\bf\Phi}'(x)={\bf\Phi}'(-x)$, ${\bf\Phi}''(x)=-x{\bf\Phi}'(x)$ and ${\bf\Phi}'(\frac{a}{b}+b)=e^{-a-\frac{b^2}2}{\bf\Phi}'(\frac{a}{b})$.

\smallskip
		The following lemma will be useful in derivation of asymptotic behavior of the early exercise. Its proof is straightforward and follows
		from monotonicity of the right-hand side of equation (\ref{h_equation}) in the $h$ variable.

\begin{lemma}\label{h_lemma}
The implicit equation
\begin{equation}
\label{h_equation}
0=1-\int_0^1{{\bf\Phi}\Big(-h\frac{1-\sqrt{1-\theta}}{\sqrt{\theta}}\Big)}d\theta+
h\int_0^1{\frac{\sqrt{1-\theta}}{\sqrt{\theta}}{\bf\Phi}'\Big(-h\frac{1-\sqrt{1-\theta}}{\sqrt{\theta}}\Big)}d\theta
\end{equation}
has the unique solution $h^*$ having its approximate value $h^* \doteq -0.638833$.
\end{lemma}
			
Notice that the first order asymptotic expansion as $t\to T$ of the early exercise boundary $S_f(t) \approx S_f(T)(1+ 0.638833\, \sigma \sqrt{T-t})$ for the American call option derived by \cite{NJD-SDH-IR-PW-1993,DS-2001} contains the same constant $-h^* \doteq 0.638833$ where $h^*$ is a solution of (\ref{h_equation}).

\begin{remark}
For the early exercise boundary function $x^*_t=x^*_t(T,r,q,\sigma^2)$ as a function of the model parameters $T,r,\sigma^2>0, q\ge 0,$ we have the following scaling property:
\begin{equation*}
x^*_t(T,r,q,\sigma^2)=x^*_{\frac{t}{T}}(1,r T,q T,\sigma^2 T).
\end{equation*}
\end{remark}
									
		\subsubsection{Geometric average}
			\label{geomC}
		
			We calculate the derivative of the European part of the expression (for the call option). Recall that $\alpha$ depends on the variable
			$x$, but $\beta$ does not depend on this variable.
			\begin{equation*}
%				\label{geom_eq_I}
				\widehat{v}_{x}^g(t,x)=e^{qt}\frac{\partial}{\partial x}\widetilde{v}^g(t,x)=
						-e^{-q(T-t)}e^{\alpha^g_{t,T}+\frac{(\beta^g_{t,T})^2}2}
						{\bf\Phi}\Big(-\frac{\alpha^g_{t,T}}{\beta^g_{t,T}}-\beta^g_{t,T}\Big)\frac{\partial\alpha^g_{t,T}}{\partial x}.
			\end{equation*}
			Now we calculate the derivative of the integral function of American-style option bonus:
			\begin{eqnarray*}
%				\label{geom_eq_II}
				\widehat{e}^g_{Ix}(t,x,u,x_u^*)=
					e^{-q(u-t)}\frac1{u}\Bigg(
							\Big(\frac{-q u+x^*_u r u+x^*_u\ln{x^*_u}+x^*_u(\beta^g_{t,u})^2}{\beta^g_{t,u}}\Big){\bf\Phi}'\Big(\frac{\ln{x^*_u}-\alpha^g_{t,u}}{\beta^g_{t,u}}\Big)\nonumber\\
					-e^{\alpha^g_{t,u}+\frac{(\beta^g_{t,u})^2}2}
						\Big(r u+\alpha^g_{t,u}+(\beta^g_{t,u})^2+1\Big){\bf\Phi}\Big(\frac{\ln{x^*_u}-\alpha^g_{t,u}}{\beta^g_{t,u}}-\beta^g_{t,u}\Big)\Bigg)\frac{\partial\alpha^g_{t,u}}{\partial x}.
			\end{eqnarray*}
			
			We want to determine the behavior of the early exercise boundary near the expiry $T$. The limit in the expression (\ref{rov_hran})
			leads to the trivial identity. By rearranging all of elements on the right side of the equation, we have an expression of order $T-t$.
			We substitute $\tau=T-t$ and $x^*_T=G$ and divide the equation by $\tau$. We have:
			\begin{equation*}
				\lim_{\tau\rightarrow0}\frac{1+\widehat{v}_{x}(T-\tau,G(1+h\sigma\sqrt{\tau}))}{\tau}=
						\lim_{\tau\rightarrow0}\frac{\partial\widehat{v}_{x}}{\partial\tau}(T-\tau,G(1+h\sigma\sqrt{\tau})).
			\end{equation*}
			According to the Lemma~\ref{geom_momenty} and Table~\ref{start_point} we have:
			\begin{equation*}
				\lim_{\tau\rightarrow0}\partial_{\tau}\alpha^g_{T-\tau,T}\,\partial_x\alpha^g_{T-\tau,T}
					+\partial^2_{x, \tau}\alpha^g_{T-\tau,T}
					=\frac1G\Big(q-\frac{q}{G}-\frac1T-\frac{\sigma^2}2\Big).
			\end{equation*}
			The only non-zero elements of the first partial limit are the elements multiplied by the cumulative distribution function.
			$$
				\lim_{\tau\rightarrow0}\frac{1+\widehat{v}_{x}(T-\tau,G(1+h\sigma\sqrt{\tau}))}{\tau}=
				\lim_{\tau\rightarrow0}\frac{\partial}{\partial\tau}\Big(1-e^{-q\tau}e^{\alpha^g_{T-\tau,T}+\frac{(\beta^g_{T-\tau,T})^2}2}
						\frac{\partial\alpha^g_{T-\tau,T}}{\partial x}\Big)=\frac{q}{G}+\frac1T.
			$$
			The second term represents the limit of the integral part divided by $\tau$. If we substitute $u=T-\tau(1-\theta)$, then we obtain 			\begin{eqnarray*}
			&&\lim_{\tau\rightarrow0}
				\frac{\int_{T-\tau}^{T}{\widehat{e}^g_{Ix}(T-\tau,G(1+h\sigma\sqrt{\tau})),u,G(1+h\sigma\sqrt{T-u}))}du}{\tau}\nonumber\\
			&&=\int_{0}^{1}{\lim_{\tau\rightarrow0}
				\widehat{e}^g_{Ix}(T-\tau,G(1+h\sigma\sqrt{\tau})),T-\tau(1-\theta),G(1+h\sigma\sqrt{\tau}\sqrt{1-\theta}))}d\theta.
			\end{eqnarray*}
			The last expression then can be simplified, using  limits (\ref{limity}), equation (\ref{transc_eq}) for the limit of early exercise
			boundary at expiry $G$, i.e. $r=\frac{q}{G}-\frac{\ln{G}}T$,
			and by calculating the limit (using L'Hospital rule) of the expression multiplied by the derivative of the cumulative density function.
			The final limit of the integral has form
			\begin{eqnarray}
				\label{int_h_geom}
				&&\lim_{\tau\rightarrow0}
				\widehat{e}^g_{Ix}(T-\tau,G(1+h\sigma\sqrt{\tau})),T-\tau(1-\theta),G(1+h\sigma\sqrt{\tau}\sqrt{1-\theta}))\nonumber\\
				&&=\Big(\frac{q}{G}+\frac1T\Big)\Bigg(-{\bf\Phi}\Big(-h\frac{1-\sqrt{1-\theta}}{\sqrt{\theta}}\Big)+
						h\frac{\sqrt{1-\theta}}{\sqrt{\theta}}{\bf\Phi}'\Big(-h\frac{1-\sqrt{1-\theta}}{\sqrt{\theta}}\Big)\Bigg).
			\end{eqnarray}
			Integrating (\ref{int_h_geom}) with respect to $\theta\in[0,1]$, putting both partial limits together,
			dividing by the nonzero constant $\frac{q}{G}+\frac1T$ and  by Lemma~\ref{h_lemma} and Table~\ref{start_point}, we finally obtain
			\begin{equation*}
				x^*_t=G(1+h^*\sigma\sqrt{T-t})+O(T-t) \quad \hbox{as}\ t\to T, \quad \hbox{where} \ h^* \doteq -0.638833.
			\end{equation*}

		\subsubsection{Arithmetic average}
		
			The derivation for the case of the arithmetic average is very similar to the geometric one.			
			We calculate the derivatives of both parts of the value function. The European part
			\begin{eqnarray*}
%				\label{arit_eq_I}
				\widehat{v}_{x}^a(t,x_t)&=&e^{qt}\frac{\partial}{\partial x}\widetilde{v}^a(t,x_t)=
						e^{-q(T-t)}{\bf\Phi}'\Big(-\frac{\alpha^a_{t,T}}{\beta^a_{t,T}}\Big)\frac{\partial\beta^a_{t,T}}{\partial x}\nonumber\\
						&&-e^{-q(T-t)+\alpha^a_{t,T}+\frac{(\beta^a_{t,T})^2}2}
						{\bf\Phi}\Big(-\frac{\alpha^a_{t,T}}{\beta^a_{t,T}}-\beta^a_{t,T}\Big)
						\Big(\frac{\partial\alpha^a_{t,T}}{\partial x}+\beta^a_{t,T}\frac{\partial\beta^a_{t,T}}{\partial x}\Big)
			\end{eqnarray*}
			and the American-style bonus
			\begin{eqnarray*}
				\widehat{e}^a_{Ix}(t,x,u,x_u^*)&=&
					e^{-q(u-t)}\Bigg(\Bigg(\Big(q+\frac1u\Big)-\Big(r+\frac1u\Big)x^*_u\Bigg)
									{\bf\Phi}'\Big(\frac{\ln{x^*_u}-\alpha^a_{t,u}}{\beta^a_{t,u}}\Big)
									\frac{\partial}{\partial x}\Big(\frac{\ln{x^*_u}-\alpha^a_{t,u}}{\beta^a_{t,u}}\Big)\nonumber\\
							&&+\Big(r+\frac1u\Big)x^*_u{\bf\Phi}'\Big(\frac{\ln{x^*_u}-\alpha^a_{t,u}}{\beta^a_{t,u}}\Big)
								\frac{\partial\beta^a_{t,u}}{\partial x}\nonumber\\
							&&-\Big(r+\frac1u\Big)e^{\alpha^a_{t,u}+\frac{(\beta^a_{t,u})^2}2}
								{\bf\Phi}\Big(\frac{\ln{x^*_u}-\alpha^a_{t,u}}{\beta^a_{t,u}}-\beta^a_{t,u}\Big)
								\Big(\frac{\partial\alpha^a_{t,u}}{\partial x}+\beta^a_{t,u}\frac{\partial\beta^a_{t,u}}{\partial x}\Big)\Bigg).
			\end{eqnarray*}
			
			The rest of derivation was performed following the same steps used in section \ref{geomC}.
			The main difference in the derivation is that for the arithmetic average also the expression  $\beta^a_{t,u}=\beta^a(t,u,x)$  depends on the variable $x$.
			Thus, according to the Lemma~\ref{arit_momenty} and Table~\ref{start_point} we need to calculate following limits
			\begin{eqnarray*}
				&&\lim_{\tau\rightarrow0}\partial_{\tau}\alpha^a_{T-\tau,T}\,\partial_x\alpha^a_{T-\tau,T}+\partial^2_{x, \tau}\alpha^a_{T-\tau,T}
												=\frac1A\Big(q-r-\frac1T-\frac{\sigma^2}2\Big),\\
				&& \lim_{\tau\rightarrow0}\partial_x\beta^a_{T-\tau,T}=\lim_{\tau\rightarrow0}\partial_x\beta^a_{T-\tau,T-\tau(1-\theta)} =
				\lim_{\tau\rightarrow0}\partial_x\beta^a_{T-\tau,T}\,\partial_x\beta^a_{T-\tau,T}+\beta^a_{T-\tau,T}\partial^2_{x, \tau}\beta^a_{T-\tau,T}=0.
			\end{eqnarray*}
			We recall that for the case of a continuous arithmetic average we have $x^*(T)=A\equiv \frac{q+\frac1T}{r+\frac1T}$.  Since we have assumed $r>q\ge0$ we obtain $A<1$ (see Table~\ref{start_point}).
			
The derivation leads to the same equation as in Lemma~\ref{h_lemma} when  multiplied by the constant $r+\frac1T$. In summary, we obtain the following approximation of the limiting behavior of the early exercise boundary near expiry also for the arithmetic average.
			\begin{equation}
				\label{aa-asymptotika}
x^*(t)=\frac{1 + q T}{1 + r T}(1+h^*\sigma\sqrt{T-t})+O(T-t), \quad \hbox{as}\ t\to T, \quad \hbox{where} \ h^*\doteq-0.638833.
			\end{equation}

\section{Transformation method for Asian call options}

	The purpose of this section is to propose an efficient numerical algorithm for determining the free boundary position $x^*_t$
	for American-style of Asian options. Construction of the algorithm is based on a solution to a nonlocal parabolic partial differential
	equation (PDE). The governing PDE is constructed for a transformed variable representing the so-called $\delta$-synthetised portfolio.
	Furthermore, we employ a front fixing method (refereed also to as Landau's fixed domain transformation) developed by \cite{RS-DS-JC-1999,DS-2001}
	for plain vanilla options as well as for a class of nonlinear Black--Scholes equations \cite{DS-2007,DS-2009}. At the end of this section we
	present numerical results and comparisons achieved by these methods to the recent method developed by \cite{MD-YKK-2006}.

	First, we recall the partial differential equation for pricing Asian options (cf. \cite{YKK-1998}). We assume the asset price dynamics follows
	a geometric Brownian with a drift $\varrho$, continuous dividend yield $q\ge 0$ and volatility $\sigma$, i.e. $d S = (\varrho -q)S dt + \sigma S dW$
	where $W$ is the standard Wiener process. If we apply It\=o's formula to the function $V=V(t,S,A)$  we obtain
	\begin{equation}
		d V = \left(
		   \frac{\partial V}{\partial t}
		+\frac{\sigma^2}{2}S^2 \frac{\partial^2 V}{\partial S^2}
		\right)dt
		+
		\frac{\partial V}{\partial S} dS
		+
		\frac{\partial V}{\partial A} dA\,.
		\label{differential-V}
	\end{equation}
	Recall that for arithmetic, geometric or weighted arithmetic averaging we have $dA/A = f(A/S, t) dt$ where the function $f=f(x,t)$ is defined as
	follows (see Tab.~\ref{AA_hodnoty}):
	\begin{equation}
		\label{general-f}
		f(x,t) = \left\{
		\begin{array}{ll}
			\frac{x^{-1}-1}{t} & \hbox{arithmetic averaging,} \\ 
			\smallskip
			\frac{-\ln x}{t} & \hbox{geometric averaging,}
			\\
			\frac{\lambda(x^{-1} -1 )}{1-e^{-\lambda t}} & \hbox{exponentially weighted arithmetic averaging.}
		\end{array} 
		\right.
	\end{equation}
	Inserting the expression $dA = A f(A/S, t) dt$ into (\ref{differential-V}) and 
	following standard arguments from the Black--Scholes theory we obtain the governing equation for pricing Asian option with averaging
	given by (\ref{general-f}) in the form:
	\begin{equation}
		\label{asian-eq}
		\frac{\partial V}{\partial t}
		+\frac{\sigma^2}{2}S^2 \frac{\partial^2 V}{\partial S^2}
		+S(r-q)\frac{\partial V}{\partial S}
		+A\, f(A/S, t) \frac{\partial V}{\partial A} - r V = 0, 
	\end{equation}
	where $0<t<T, \ S, A>0$ (see e.g. \cite{MD-YKK-2006,YKK-1998}). For the Asian call option the above equation is subject to the terminal pay-off condition
	$V(T,S,A) = \max(S- A,0), \quad S,A>0$. 
	It is well known (see e.g. \cite{YKK-1998,MD-YKK-2006}) that for Asian options with floating strike we can perform dimension reduction by introducing the
	following similarity variable:
	\[
		x=A/S, \qquad W(x,\tau) = V(t,S,A)/A
	\]
	where $\tau=T-t$. It is straightforward to verify that $V(t,S,A) =  A\, W(A/S, T-t)$ is a solution of (\ref{asian-eq}) iff $W=W(x,\tau)$ is a solution to the following parabolic PDE:
	\begin{equation}
		\label{asian-redeq}
		\frac{\partial W}{\partial \tau}
		- \frac{\sigma^2}{2}\frac{\partial}{\partial x}\left(x^2 \frac{\partial W}{\partial x}\right)
		+ (r-q)x\frac{\partial W}{\partial x}
		-f(x,T-\tau)\left( W+x\frac{\partial W}{\partial x}\right)+rW=0, 
	\end{equation}
	where $x>0$ and $0<\tau<T$.  The initial condition for $W$ immediately follows from the terminal pay-off diagram for the call option, i.e. 
	$
		W(x,0) = \max(x^{-1}-1,0).
	$

	\subsection{American-style of Asian call options}

		Following \cite{MD-YKK-2006} the set 
		$
			{\mathcal E} = \{ (t,S,A) \in [0,T]\times [0,\infty)\times[0,\infty),\ 
			V(t,S,A) = V(T,S,A)\}
		$
		is the exercise region for American-style of Asian call options. 
		In the case of a call option this region can be described by the early exercise boundary function $S_f=S_f(t,A)$ such that 
		${\mathcal E} = \{ (t,S,A) \in [0,T]\times[0,\infty)\times[0,\infty),\ 
		S\ge S_f(t,A)\}$. 
		For American-style of an Asian call option we have to impose a homogeneous Dirichlet boundary condition $V(t,0,A)=0$.
		According to \cite{MD-YKK-2006} the $C^1$ continuity condition at the point  $(t,S_f(t,A),A)$ of a contact of a solution $V$ with
		its pay-off diagram implies the following boundary condition at the free boundary position $S_f(t,A)$: 
		\begin{equation}
			\label{asian-bc}
			\frac{\partial  V}{\partial S}(t,S_f(t,A),A)=1,\quad 
			V(t,S_f(t,A),A)=S_f(t,A) - A,
		\end{equation}
		for any $A>0$ and $0<t<T$. It is important to emphasize that the free boundary function $S_f$ can be also reduced to a function of one
		variable by introducing a new state function $x^*_t$ as follows:
		\[
			S_f(t,A) = A/x^*_t.
		\]
		The function  $t\mapsto x^*_t$ is a free boundary function for the transformed state variable $x=A/S$. For American-style of Asian call
		options the spatial domain for the reduced equation (\ref{asian-redeq}) is given by
		$
			1/\ro(\tau) < x < \infty,\ \ \tau\in(0,T),\quad \hbox{where}\ \ro(\tau)=1/x^*_{T-\tau}\,.
		$
		Taking into account boundary conditions (\ref{asian-bc}) for the option price $V$ we end up with corresponding boundary conditions for
		the function $W$:
		\begin{equation}
			\label{asian-redbc}
			W(\infty,\tau) =0, \qquad W(x,\tau)=\frac{1}{x}-1,\ \ \frac{\partial W}{\partial x} (x,\tau) = - \frac{1}{x^2} \ \ \hbox{at}\ \ x=\frac{1}{\ro(\tau)}
		\end{equation}
		for any $0<\tau<T$ and the initial condition 
		\begin{equation}
			\label{asian-redic}
			W(x,0) =\max(x^{-1}-1,0) \quad \hbox{for any}\  x>0. 
		\end{equation}

	\subsection{Fixed domain transformation}

		In order to apply the Landau fixed domain transformation for the  free boundary problem (\ref{asian-redeq}), (\ref{asian-redbc}),
		(\ref{asian-redic}) we introduce a new state variable $\xi$ and an auxiliary function $\Pi=\Pi(\xi,\tau)$  representing a synthetic
		portfolio. They are defined as follows:
		\begin{equation*}
%			\label{asian-portf}
			\xi = \ln\left( \ro(\tau) x \right),\qquad 
			\Pi(\xi,\tau)=W(x,\tau)+ x\frac{\partial W}{\partial x}(x,\tau)\,.
		\end{equation*}
		Clearly, $x\in (\ro(\tau)^{-1},\infty)$ iff $\xi\in (0,\infty)$ for $\tau\in(0,T)$. The value $\xi=\infty$ of the transformed variable
		corresponds to the value $x=\infty$, i.e. $S=0$ when expressed in the original variable. On the other hand, the value $\xi=0$ corresponds
		to the free boundary position $x=x^*_t$, i.e. $S= S_f(t,A)$. After straightforward calculations we conclude that the function
		$\Pi=\Pi(\xi,\tau)$ is a solution to the following parabolic PDE:
		\begin{equation*}
%			\label{asian-fixedeq}
			\frac{\partial \Pi}{\partial \tau}
			+a(\xi,\tau)\frac{\partial \Pi}{\partial \xi} 
			-\frac{\sigma^2}{2}\frac{\partial^2 \Pi}{\partial \xi^2}
			+b(\xi,\tau) \Pi=0,  
		\end{equation*}
		where the term $a(\xi,\tau)$ depends on the free boundary position $\varrho$. The terms $a,b$ are given given by
		\begin{equation}
			a(\xi,\tau)=\frac{\dot{\varrho}(\tau)}{\varrho(\tau)}
			+ r - q -\frac{\sigma^2}{2}-f(e^{\xi}/\varrho(\tau), T-\tau),\ 
			b(\xi,\tau) = r- \frac{\partial}{\partial x}\left(x f(x,T-\tau)\right)\bigg|_{x=\frac{e^\xi}{\varrho(\tau)}}.
			\label{cleny-ab}
		\end{equation}
		Notice that 
		$
			b(\xi,\tau) = r + 1/(T-\tau)
		$
		in the case of arithmetic averaging, i.e. $f(x,t) = (x^{-1} -1 )/t$. 
		
The initial condition for the solution $\Pi$ can be determined from (\ref{asian-redic})
		\[
			\Pi(\xi,0)    =  \left\{ \begin{array}{ll}
					-1 & \mbox{$\xi<\ln \varrho(0)$}, \\
					0 & \mbox{$\xi>\ln \varrho(0)$}.
					\end{array} \right.
		\]
		Since $\partial_x W(x,\tau) = -\frac{1}{x^2}$ and $W(x,\tau)=\frac{1}{x}-1$ for $x=\frac{1}{\ro(\tau)}$ and $W(\infty,\tau) = 0$ we
		conclude the Dirichlet boundary conditions for the transformed function $\Pi(\xi,\tau)$
		\[
			\Pi(0,\tau) = - 1, \qquad \Pi(\infty,\tau)=0.
		\]
		It remains to determine an algebraic constraint between the free boundary function $\ro(\tau)$ and the solution $\Pi$. Similarly as
		in the case of a linear or nonlinear Black--Scholes equation (cf. \cite{DS-2007}) we obtain, by differentiation the condition
		$W(\frac{1}{\varrho(\tau)},\tau)=\varrho(\tau)-1$ with respect to $\tau,$ the following identity:
		\[
			\frac{d\varrho}{d\tau}(\tau)=\frac{\partial W}{\partial x}(\varrho(\tau)^{-1},\tau) \left(-\varrho(\tau)^{-2}\right)  \frac{d\varrho}{d\tau}(\tau)
					+\frac{\partial W}{\partial \tau}(\varrho(\tau)^{-1},\tau).
		\]
		Since $\partial_x W(\varrho(\tau)^{-1},\tau)=-\varrho(\tau)^{2}$ we have 
		$\frac{\partial W}{\partial \tau}(x ,\tau)=0$ at $x=\ro(\tau)$. 
		Assuming continuity of the function  $\Pi(\xi,\tau)$ and its derivative $\Pi_\xi(\xi,\tau)$ up to the boundary $\xi=0$ we obtain
		\[
			x^2\frac{\partial^2 W}{\partial x^2}(x,\tau)
			\to
			\frac{\partial \Pi}{\partial \xi}(0,\tau) +2\varrho(\tau),
			\quad 
			x\frac{\partial W}{\partial x}(x,\tau)\to - \varrho(\tau) \quad \hbox{as}
			\ x\to \varrho(\tau)^{-1}.
		\]
		Passing to the limit $x\to \varrho(\tau)^{-1}$ in (\ref{asian-redeq}) we end up with the algebraic equation 
		\begin{equation}
			\label{general-rhoeq}
			q\varrho(\tau) -r  + f(\varrho(\tau)^{-1}, T-\tau)= \frac{\sigma^2}{2}\frac{\partial \Pi}{\partial\xi}(0,\tau)
		\end{equation}
		for the free boundary position $\varrho(\tau)$ where $\tau\in(0,T]$. Notice that, in the case of arithmetic averaging where
		$f(\varrho(\tau)^{-1}, T-\tau) = (\varrho(\tau)-1)/(T-\tau)$, we can derive an explicit expression for the free boundary position $\ro(\tau)$
		\begin{equation*}
%			\label{aa-rhoeq}
			\varrho(\tau)
			=\frac{1+r(T-\tau) + \frac{\sigma^2}{2}(T-\tau)\frac{\partial \Pi}{\partial        \xi}(0,\tau)}{1+q(T-\tau)}, \quad 0<\tau<T,
		\end{equation*}
		as a function of the derivative $\partial_\xi\Pi(0,\tau)$ evaluated at $\xi=0$. The value $\varrho(0)$ can be deduced from
		Theorem~\ref{st_p}. For the arithmetic averaging we have (see also \cite{MD-YKK-2006}) the following expression:
		\[
			\ro(0)=\max\left(\frac{1+r T}{1+q T},1\right).
		\]

In summary, we derived the following nonlocal parabolic equation for the synthesized portfolio $\Pi(\xi,\tau)$: 
		\begin{eqnarray}
			\label{asian-system}
			&&\frac{\partial \Pi}{\partial
					\tau}+a(\xi,\tau)\frac{\partial \Pi}{\partial \xi}
					-\frac{\sigma^2}{2}\frac{\partial^2 \Pi}{\partial
					\xi^2}+ b(\xi,\tau) \Pi=0,\quad 0< \tau <T,\ 
					\xi>0,\nonumber \\
			&&\hskip-1truecm\hbox{with an algebraic constraint}\nonumber \\
			&& 
			q\varrho(\tau) -r  + f(\varrho(\tau)^{-1}, T-\tau)= \frac{\sigma^2}{2}\frac{\partial \Pi}{\partial\xi}(0,\tau), \ \ 0<\tau<T, \nonumber \\
			&&\hskip-1truecm\hbox{subject to the boundary and initial conditions} \\
			&& \Pi(0,\tau) =-1, \qquad \Pi(\infty,\tau)=0, \nonumber \\
			&& \Pi(\xi,0)   =  \left\{ 
			\begin{array}{rl}
				-1 & \mbox{for $\xi<\ln(\varrho(0))$}, \\
				 0 & \mbox{for $\xi>\ln(\varrho(0))$},
			\end{array} \right. \nonumber \\
			&&\hskip-1truecm\hbox{where $a(\xi,\tau)$ and $b(\xi,\tau)$ are given by (\ref{cleny-ab}),} \nonumber\\ 
			&&\hskip-1truecm\hbox{and the starting point $\varrho(0)=1/x^*_T$ is given by Theorem~\ref{st_p}.} \nonumber 
		\end{eqnarray}

		\subsubsection{An equivalent form of the equation for the free boundary}
		
			Although equation (\ref{general-rhoeq}) provides an algebraic formula for the free boundary position $\varrho(\tau)$
			in terms of the derivative $\partial_\xi\Pi(0,\tau)$ such an expression is not quite suitable for construction of a robust
			numerical approximation scheme. The reason is that any small inaccuracy in approximation of the value $\partial_\xi\Pi(0,\tau)$
			is transferred in to the entire computational domain $\xi\in(0,\infty)$ making thus a numerical scheme very sensitive to the value
			of the derivative of a solution evaluated in one point $\xi=0$. In what follows, we present an equivalent equation for the free
			boundary position $\varrho(\tau)$ which is more robust from the numerical approximation point of view. 

			Integrating the governing equation (\ref{asian-system}) with respect to $\xi\in(0,\infty)$ yields
			\[
				\frac{d}{d\tau} \int_0^\infty \hskip-2truemm\Pi d\xi 
				+ \int_0^\infty  \hskip-2truemm a(\xi,\tau)\frac{\partial \Pi}{\partial \xi}d\xi
				-\frac{\sigma^2}{2}\int_0^\infty \frac{\partial^2 \Pi}{\partial\xi^2} d\xi
				+\int_0^\infty  \hskip-2truemm b(\xi,\tau) \Pi d\xi=0.
			\]
Now, taking into account the boundary conditions $\Pi(0,\tau)=-1, \Pi(\infty,\tau)=0$, and consequently
$\partial_\xi\Pi(\infty,\tau)=0$ we obtain, by applying condition (\ref{general-rhoeq}), the following differential equation:
\begin{eqnarray*}
\frac{d}{d\tau}\left(\ln\varrho(\tau) + \int_0^\infty \hskip-2truemm\Pi(\xi,\tau) d\xi \right) 
+ q\varrho(\tau) - q - \frac{\sigma^2}{2}
+ \int_0^\infty \hskip-2truemm\left[r - f\left(\frac{e^\xi}{\varrho(\tau)}, T-\tau\right)\right] \Pi(\xi,\tau) d\xi=0.
\end{eqnarray*}
In the case of arithmetic averaging where $f(x,t) = (x^{-1}-1)/t$ we obtain
\begin{equation}
\label{aa-inteq}
\frac{d}{d\tau}\left(\ln\varrho(\tau) + \int_0^\infty \hskip-2truemm\Pi(\xi,\tau) d\xi \right)
+   q\varrho(\tau) - q - \frac{\sigma^2}{2} 
+ \int_0^\infty \left[r- \frac{\varrho(\tau) e^{-\xi} -1 }{T-\tau}\right]  \Pi(\xi,\tau) d\xi=0.
\end{equation}

	\subsection{A numerical approximation operator splitting scheme}
	
		Our numerical approximation scheme is based on a solution to the transformed system (\ref{asian-system}). For the sake of simplicity,
		the scheme will be derived for the case of arithmetically averaged Asian call option. Derivation of the scheme for geometric or weighted
		arithmetic averaging is similar and therefore omitted. 

		We restrict the spatial domain $\xi\in (0,\infty)$ to a finite interval of values $\xi\in (0,L)$ where $L>0$ is sufficiently large.
		For practical purposes it sufficient to take $L\approx 2$. Let $k>0$ denote by the time step, $k=T/m$, and, by $h=L/n>0$ the spatial step. Here 
		$m, n\in {\mathbb N}$ denote the number of time and space discretization steps, resp. We denote by $\Pi^j=\Pi^j(\xi)$
		the time discretization  of $\Pi( \xi, \tau_j)$ and $\ro^j \approx \ro(\tau_j)$ where $\tau_j = j k$. By  $\Pi^j_i$ we shall denote
		the full space--time approximation for the value $\Pi(\xi_i, \tau_j)$. Then for the Euler backward in time finite difference approximation
		of equation (\ref{asian-system}) we have
		\begin{equation*}
%			\label{asian-timediscret}
			\frac{\Pi^j-\Pi^{j-1}}{k} +  c^j \frac{\partial \Pi^j}{\partial \xi} 
			- \left( \frac{\sigma^2}{2} + \frac{\varrho^j e^{-\xi}-1}{T- \tau_j} \right) \frac{\partial \Pi^j}{\partial \xi}
			-\frac{\sigma^2}{2} \frac{\partial^2 \Pi^j}{\partial^2 \xi} 
			+\left(r +\frac{1}{T- \tau_j}\right) \Pi^j = 0
		\end{equation*}
		where $c^j$ is an approximation of the value $c(\tau_j)$ where the 
		$c(\tau)= \frac{\dot\ro(\tau)}{\ro(\tau)} + r - q $. The solution $\Pi^j=\Pi^j(x)$ is subject to Dirichlet boundary conditions at $\xi=0$
		and $\xi=L$. We set $\Pi^0(\xi)=\Pi(\xi,0)$ (see (\ref{asian-system})). In what follows, we make use of the time step operator splitting method.
		We split the above problem into  a convection part and a diffusive part by introducing  an auxiliary intermediate step $\Pi^{j-\onehalf}$:

		({\it Convective part})
		\begin{equation}
			\label{asian-convective}		
			\frac{\Pi^{j-\onehalf}-\Pi^{j-1}}{k} + c^j  \partial_x\Pi^{j-\onehalf} = 0\,,
		\end{equation}

		({\it Diffusive part})
		\begin{equation}
			\label{asian-diffusion}
			\frac{\Pi^{j}-\Pi^{j-\onehalf}}{k}  
			- \left( \frac{\sigma^2}{2} + \frac{\varrho^j e^{-\xi}-1}{T- \tau_j} \right) \frac{\partial \Pi^j}{\partial \xi}
			-\frac{\sigma^2}{2} \frac{\partial^2 \Pi^j}{\partial^2 \xi} 
			+\left(r +\frac{1}{T- \tau_j}\right) \Pi^j = 0.
		\end{equation}
		Similarly as in \cite{DS-2007} we shall approximate the convective part by the explicit solution to the transport equation
		$\partial_\tau\tilde\Pi + c(\tau) \partial_\xi\tilde\Pi =0$ for $\xi>0$ and $\tau\in(\tau_{j-1},\tau_j]$ subject to the boundary
		condition $\tilde\Pi(0,\tau) =-1$ and the initial condition $\tilde\Pi(\xi,\tau_{j-1})=\Pi^{j-1}(\xi)$. It is known that the free
		boundary function $\ro(\tau)$ need not be monotonically increasing (see e.g. \cite{MD-YKK-2006,DS-2009} or \cite{ATH-PLJ-2000}).
		Therefore depending whether the value of $c(\tau)$ is positive or negative the boundary
		condition $\tilde\Pi(0,\tau) =-1$ at $\xi=0$ is either in--flowing ($c(\tau)>0$)  or out--flowing ($c(\tau)<0$). Hence the boundary
		condition $\Pi(0,\tau) = -1$ can be prescribed only if $c(\tau_j)\ge 0$. Let us denote by $C(\tau)$ the primitive function to $c(\tau)$,
		i.e. $C(\tau)=\ln \ro(\tau) + (r-q)\tau$. Solving the transport equation  $\partial_\tau\tilde\Pi + c(\tau) \partial_\xi\tilde\Pi =0$
		for $\tau\in [\tau_{j-1},\tau_j]$ subject to the initial condition $\Pi(\xi, \tau_{j-1}) = \Pi^{j-1}(\xi)$ we obtain:
		$\tilde\Pi(\xi,\tau)=\Pi^{j-1}(\xi-C(\tau)+C(\tau_{j-1}))$ if $\xi-C(\tau)+C(\tau_{j-1})>0$ and  $\tilde\Pi(\xi,\tau)=-1$ otherwise.
		Hence the full time-space approximation of the  half-step solution $\Pi^{j-\onehalf}_i$ can be obtained from the formula
		\begin{equation}
			\label{asian-convective-discrete}
			\Pi^{j-\onehalf}_i=\left\{ 
			\begin{matrix}
				\Pi^{j-1}(\eta_i), \hfill & \quad \hbox{if } \eta_i=\xi_i- \ln\ro^j + \ln\ro^{j-1} - (r-q)k>0\,, \hfill\cr 
				-1, \hfill & \quad \hbox{otherwise.} \hfill
			\end{matrix}
			\right.
		\end{equation}
		In order to compute the value $\Pi^{j-1}(\eta_i)$ we make use of a linear interpolation between discrete values
		$\Pi^{j-1}_i, i=0,1, ..., n$.

		Using central finite differences for approximation of the derivative $\partial_x\Pi^j$ we can approximate the diffusive part of
		a solution of (\ref{asian-diffusion}) as follows:
		\begin{eqnarray*}
			\frac{\Pi_i^j - \Pi_i^{j-\onehalf}}{k} &+& \left(r + \frac{1}{T- \tau_j}\right) \Pi_i^j
			 \\
			&-& \left( \frac{\sigma^2}{2}  + \frac{\varrho^j e^{-\xi_i}-1}{T- \tau_j} \right)
			\frac{\Pi_{i+1}^j-\Pi_{i-1}^j}{2h}
			- 
			\frac{\sigma^2}{2} 
			\frac{ \Pi_{i+1}^j - 2 \Pi_{i}^j + \Pi_{i-1}^j}{h^2}
			 =0\,.\nonumber
		\end{eqnarray*}
		Therefore the vector of discrete values $\Pi^j=\{\Pi_i^j, i=1,2, ..., n\}$ at the time level  $j\in\{1,2, ..., m\}$ is a solution
		of a tridiagonal system of linear equations
		\begin{equation}
			\label{asian-eq-tridiag}
			\alpha_i^j \Pi_{i-1}^j + \beta_i^j \Pi_{i}^j + \gamma_i^j \Pi_{i+1}^j = \Pi_i^{j-\onehalf},
\quad \hbox{for}\ \  i=1,2, ..., n,  \quad \hbox{where}
		\end{equation}
		\[
			\alpha_i^j(\ro^j)
			= 
			- \frac{k}{2h^2} \sigma^2
			+ \frac{k}{2h} \left( \frac{\sigma^2}{2} + \frac{\varrho^j e^{-\xi_i}-1}{T- \tau_j}\right),\ \ 
			\gamma_i^j(\ro^j) = 
			- \frac{k}{2h^2} \sigma^2
			- \frac{k}{2h} \left( \frac{\sigma^2}{2} + \frac{\varrho^j e^{-\xi_i}-1}{T- \tau_j}\right), \nonumber
		\]
		\begin{eqnarray}
			\label{asian-abc}
			&& \beta_i^j(\ro^j)=  1 
			+ \left(r + \frac{1}{T-\tau_j}\right) k - (\alpha_i^j +\gamma_i^j)\,.
		\end{eqnarray}
		The initial and boundary conditions at $\tau=0$ and $x=0, L,$ can be approximated as follows:
		\[
			\Pi_i^0 = \left\{
			\begin{array}{lll}
				-1 & \ \ \hbox{for} \ \xi_i <\ln\left((1+r T)/(1+ q T)\right),\hfil
				\\
				\ \ \ 0 & \ \ \hbox{for} \ \xi_i \ge \ln\left({(1+r T)/(1+ q T)}\right), \hfil
			\end{array}
			\right.
		\]
		for $i=0,1, ..., n,$ and $\Pi_0^j = -1, \ \Pi_n^j = 0$ for $j=1,...,m$.

		Finally, we employ the differential equation (\ref{aa-inteq}) to determine the free boundary position $\ro$. Taking the Euler finite
		difference approximation of $\frac{d}{d\tau}\left(\ln\varrho + \int_0^\infty \Pi  d\xi\right)$ we obtain

		({\it Algebraic part})
		\begin{equation}
			\label{asian-eq-ro}
			\ln \varrho^j = \ln \varrho^{j-1} + I_0(\Pi^{j-1})
			- I_0(\Pi^{j}) + k \left( q +  \frac{\sigma^2}{2} - q\varrho^{j-1} -  I_1(\varrho^{j-1}, \Pi^j) \right)
		\end{equation}
		where $I_0(\Pi)$ stands for numerical trapezoid quadrature of the integral $\int_0^\infty  \Pi(\xi) d\xi$ whereas
		$I_1(\varrho^{j-1},\Pi)$ is a trapezoid quadrature of the integral
		$\int_0^\infty \left(r- \frac{\varrho^{j-1} e^{-\xi} -1 }{T-\tau_j}\right)  \Pi(\xi) d\xi$.

		We formally rewrite discrete equations (\ref{asian-convective-discrete}), (\ref{asian-eq-tridiag}) and (\ref{asian-eq-ro})
		in the operator form:
		\begin{equation}
			\label{asian-abstract}
			\ro^j = {\mathcal F}(\Pi^j),\qquad
			\Pi^{j-\onehalf} ={\mathcal T}(\ro^j),\qquad
			{\mathcal A}(\ro^j) \Pi^j = \Pi^{j-\onehalf},
		\end{equation}
		where 
		$\ln {\mathcal F}(\Pi^j)$ is the right-hand side of  equation (\ref{asian-eq-ro}),
		${\mathcal T}(\ro^j)$ is the transport equation solver given by 
		the right-hand side of (\ref{asian-convective-discrete}) and
		${\mathcal A}={\mathcal A}(\ro^j)$ is a tridiagonal matrix with coefficients given 
		by (\ref{asian-abc}). The system (\ref{asian-abstract}) can be approximately solved by means of successive iterations
		procedure. We define, for $j\ge 1,$  $\Pi^{j,0} = \Pi^{j-1}, \ro^{j,0} = \ro^{j-1}$. Then the 
		$(p+1)$-th approximation of $\Pi^j$ and $\ro^j$ is obtained as a solution to the system:
		\begin{eqnarray}
			\ro^{j,p+1} &&= {\mathcal F}(\Pi^{j,p}),\nonumber\\
			\Pi^{j-\onehalf, p+1} &&={\mathcal T}(\ro^{j,p+1}),\label{asian-abstract-iter}\\
			{\mathcal A}( \ro^{j,p+1}) \Pi^{j,p+1} &&= \Pi^{j-\onehalf, p+1}\,.\nonumber
		\end{eqnarray}
		Supposing the sequence of approximate discretized solutions $\{(\Pi^{j,p}, \ro^{j,p}) \}_{p=1}^\infty$ converges to 
		the limiting value $(\Pi^{j,\infty}, \ro^{j,\infty})$ as $p\to\infty$ then this limit is a solution to a nonlinear  system of
		equations (\ref{asian-abstract}) at the time level $j$ and we can  proceed by computing  the approximate solution in the next time
		level $j+1$.

	\subsection{Computational examples of the free boundary approximation}

		Finally we present several computational examples of application of the numerical approximation scheme (\ref{asian-abstract-iter})
		for the solution $\Pi(\xi,\tau)$ and the free boundary position $\varrho(\tau)$ of (\ref{asian-system}). We consider American-style
		of Asian arithmetically averaged floating strike call options.

\begin{figure}
			\begin{center}
				\includegraphics[width=0.35\textwidth]{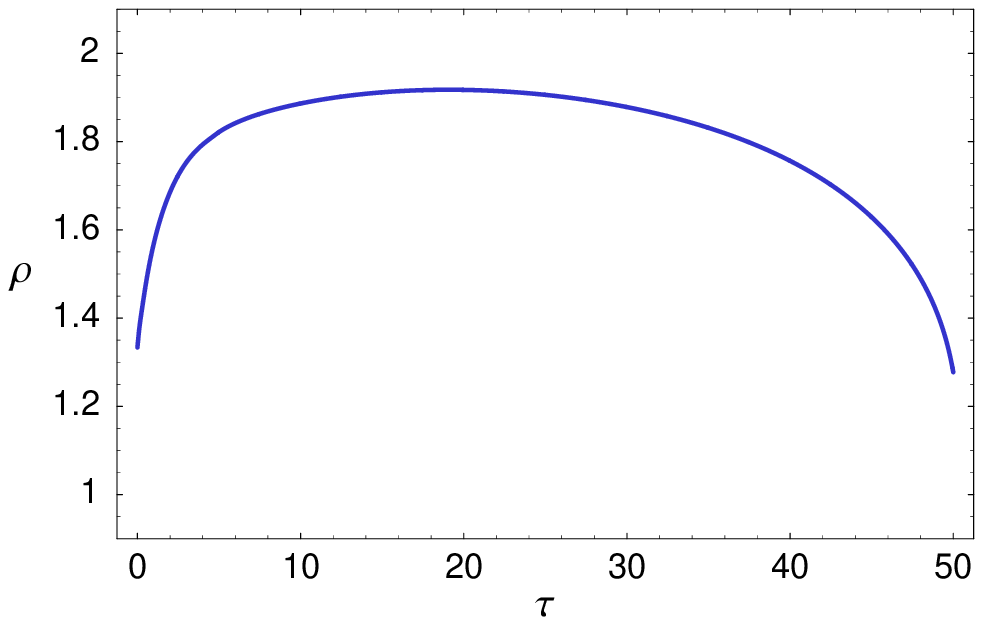}
				\includegraphics[width=0.35\textwidth]{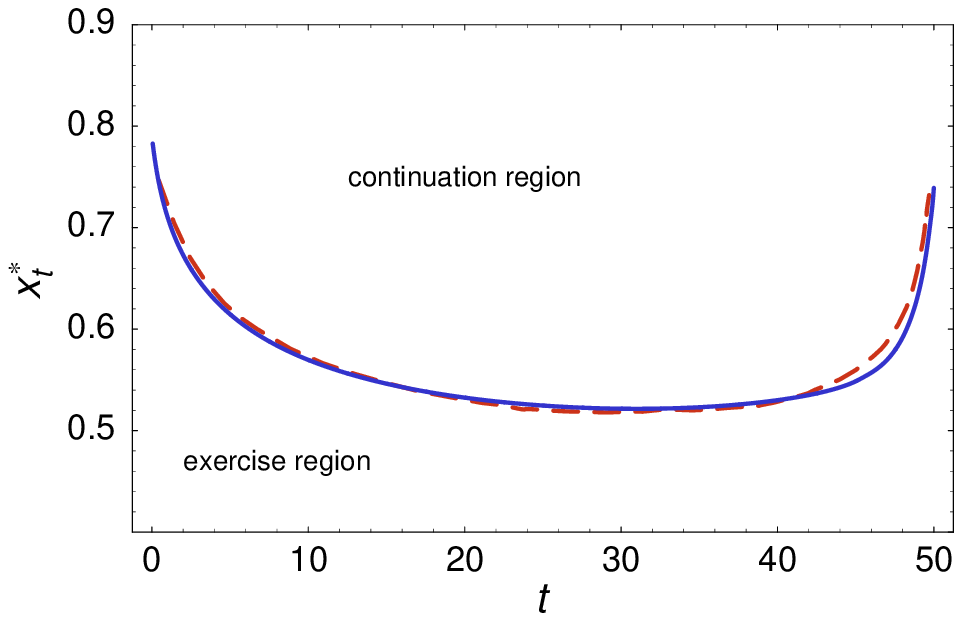}
			\end{center}
			\caption{\small 
			The function $\varrho(\tau)$ (left). A comparison of the free boundary position  $x^*_t=1/\varrho(T-t)$ (right) obtained by
			our method (solid curve) and that of the projected successive over relaxation algorithm by Dai and Kwok  (dashed curve).}
			\label{asian-fig-1}
		\end{figure}

		In Fig.~\ref{asian-fig-1} we show the behavior of the early exercise boundary function $\ro(\tau)$ and the function $x^*_t=1/\ro(T-t)$.
		In this numerical experiment we chose $r = 0.06, q= 0.04, \sigma =0.2$ and very long expiration time $T=50$ years. These parameters
		correspond to the example presented by \cite{MD-YKK-2006}. As far as other numerical parameters are concerned, we chose the mesh of
		$n=200$ spatial grid points and we have chosen the  number of time steps $m = 10^5$ in order to achieve very fine time stepping
		corresponding to 260 minutes between consecutive time steps when expressed in the original time scale of the problem.

		\begin{figure}
			\begin{center}
				\includegraphics[width=0.32\textwidth]{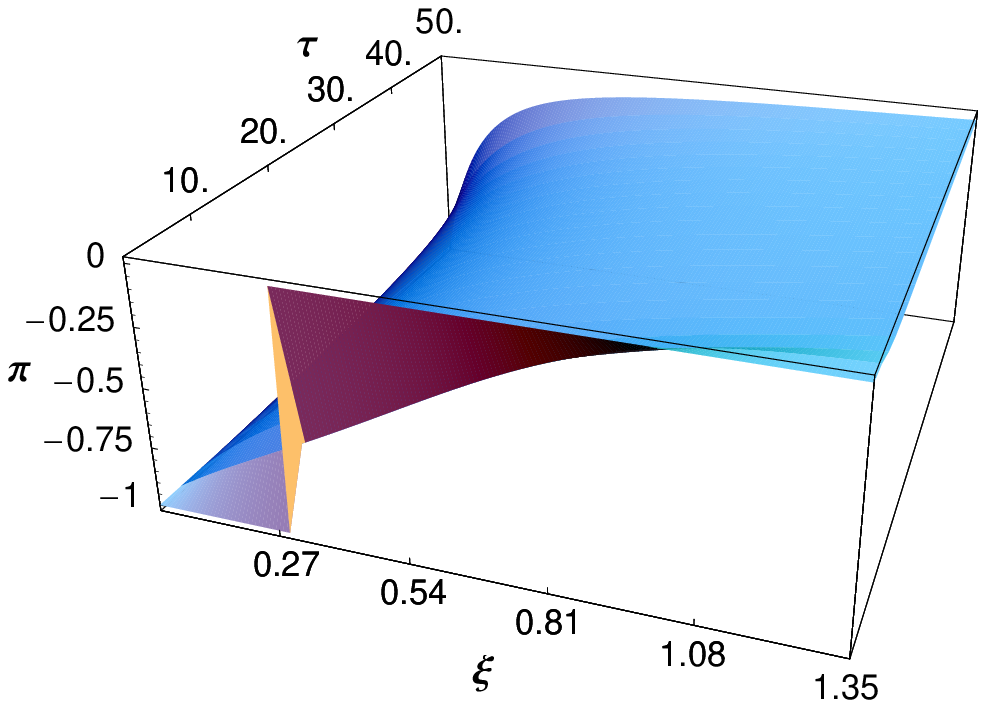}
				\includegraphics[width=0.3\textwidth]{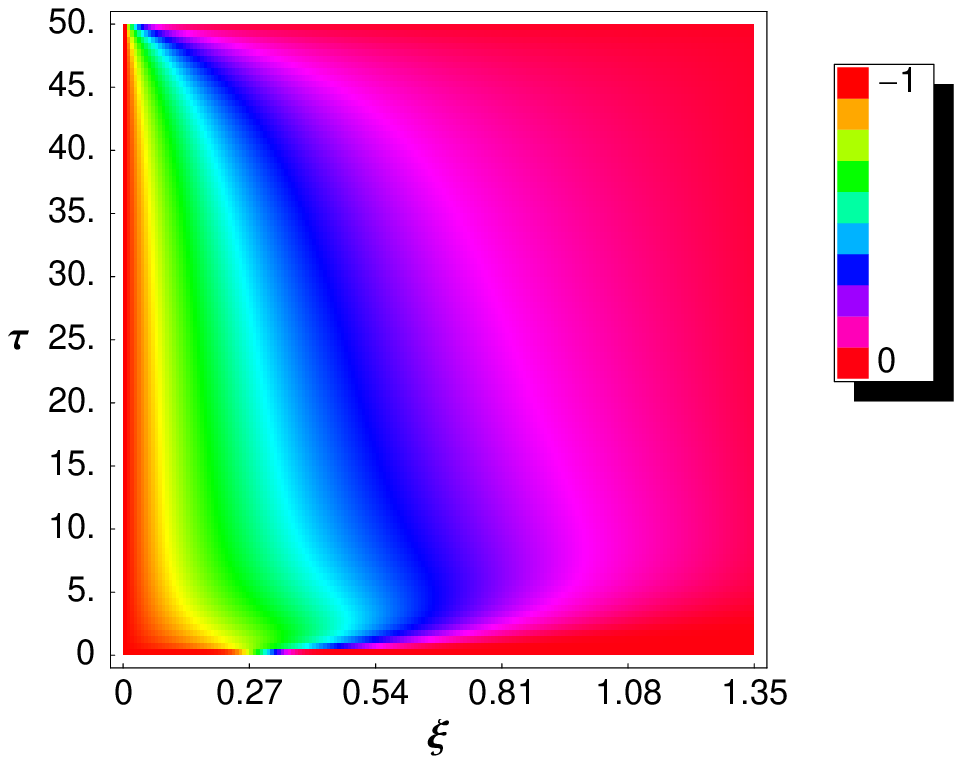}
				\includegraphics[width=0.32\textwidth]{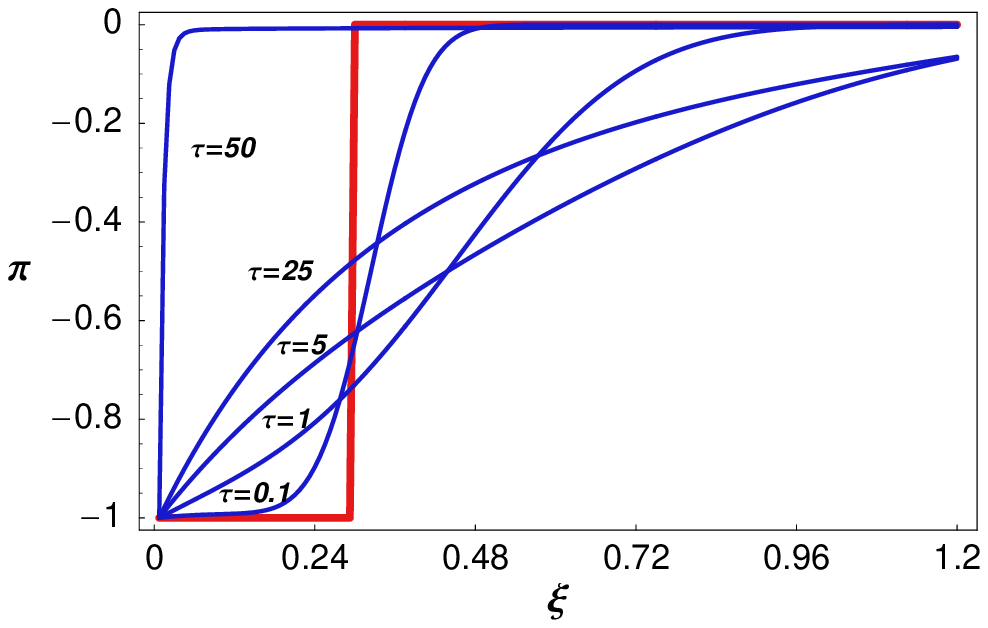}
			\end{center}
			\caption{\small 
			A 3D plot (left) and contour plot (right) of the function $\Pi(\xi,\tau)$. Profiles of the function $\Pi(\xi,\tau)$ for various
			times $\tau\in[0,T]$.}
			\label{asian-fig-2}
		\end{figure}

		In Fig.~\ref{asian-fig-2} we can see the behavior of the transformed function $\Pi$ in both 3D as well as contour plot perspectives.
		We also plot the initial condition $\Pi(\xi,0)$ and five time steps of the function $\xi\mapsto \Pi(\xi,\tau_j)$ for
		$\tau_j= 0.1,1,5,25,50$. 

\begin{figure}
			\begin{center}
				\includegraphics[width=0.35\textwidth]{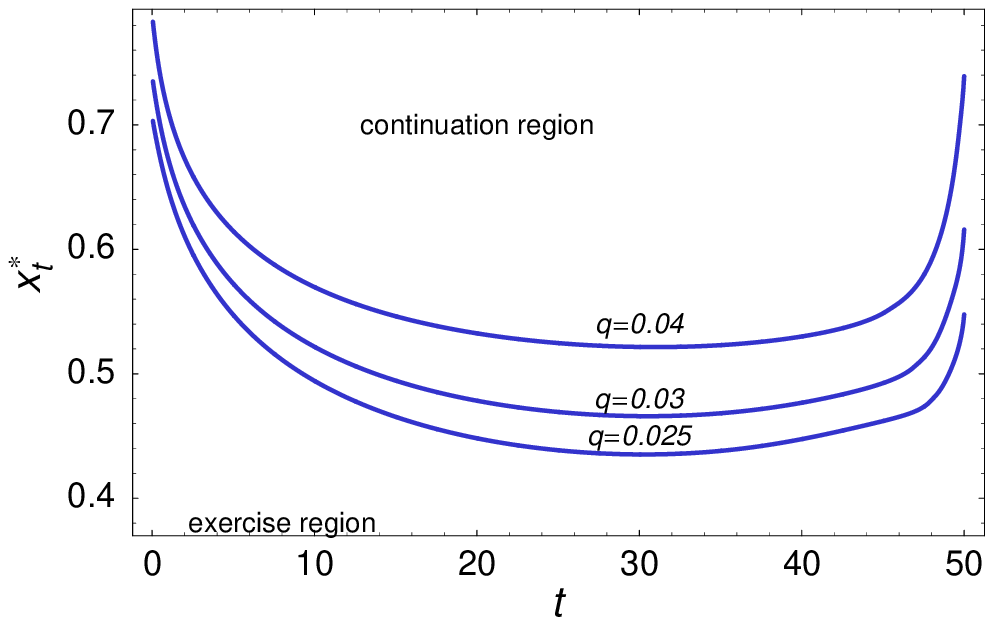}
				\includegraphics[width=0.35\textwidth]{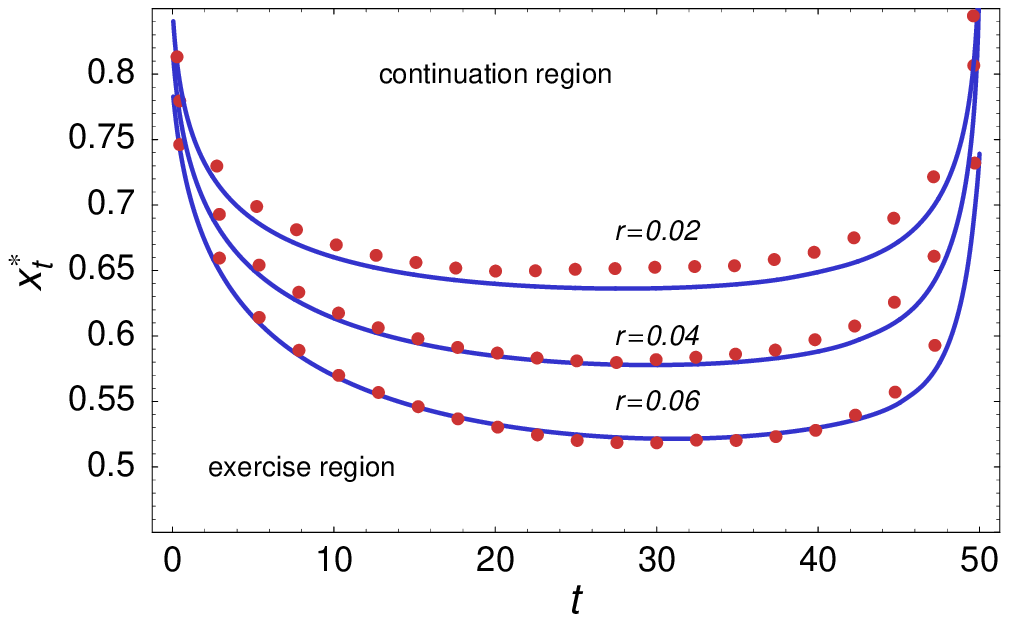}
			\end{center}
			\caption{\small 
			A comparison of the free boundary position $x^*_t$ for various dividend yield rates $q=0.04, 0.03, 0.025$ and fixed interest
			rate $r=0.06$ (left). Comparison of $x^*_t$ for various interest rates $r=0.06, 0.04, 0.02$ and fixed dividend yield $q=0.04$.
			Dots represents the solution obtained by Dai and Kwok (right).}
			\label{asian-fig-5}
		\end{figure}

		A comparison of early exercise boundary profiles with respect to varying interest rates $r$ and dividend yields $q$ is shown in
		Fig.~\ref{asian-fig-5}. A comparison of the free boundary position $x^*_t=1/\varrho(T-t)$ obtained by our method (solid curve) and
		that of the projected successive over relaxation algorithm by \cite{MD-YKK-2006} (dashed curve) for different values of the interest
		rate $r$ is shown in Fig.~\ref{asian-fig-5} (right). The algorithm of \cite{MD-YKK-2006} is based on a numerical solution to the
		variational inequality for the function $W=W(x,\tau)$ which a solution to (\ref{asian-redeq}) in the continuation region and it is smoothly
		pasted to its pay-off diagram (\ref{asian-redbc}). It is clear that our method and that of \cite{MD-YKK-2006} give almost the same
		results. A quantitative comparison of both methods is given in Table~\ref{tab-comparison} for model parameters  $T=50,\sigma=0.2, q=0.04$
		and various interest rates $r=0.02, 0.04, 0.06$. We evaluated discrete $L^\infty(0,T)$ and $L^1(0,T)$ norms of the difference $x^{*,trans}_t - x^{*,psor}_t$ between the numerical solution $x^{*,trans}_t, t\in [0,T],$ obtained by our method and that of \cite{MD-YKK-2006} denoted by  $x^{*,psor}_t$. We also show the minimal value 
$\min_{t\in [0,T]} x^{*,trans}_t$ of the early exercise boundary. 

		\begin{table}[ht]
			\scriptsize
			\begin{center}
				
				\caption{\small \label{tab-comparison} Comparison of PSOR and our transformation method for $T=50,\sigma=0.2, q=0.04$.}
				\begin{tabular}
					{p{119pt}||p{48pt}|p{48pt}|p{48pt}}
				~& 
					$r=0.06$& 
					$r=0.04$& 
					$r=0.02$ \\
					\hline\hline
					$\| x^{*,trans}_t - x^{*,psor}_t \|_\infty$& 
					0.09769& 
					0.03535& 
					0.05359 
					\\
					\hline
					$\| x^{*,trans}_t - x^{*,psor}_t \|_1$& 
					0.00503& 
					0.00745& 
					0.01437 \\
					\hline
					$\min x^{*,trans}_t$& 
					0.52150& 
					0.57780& 
					0.63619 \\
				\end{tabular}
			\end{center}
		\end{table}

\begin{figure}
			\begin{center}
				\includegraphics[width=0.32\textwidth]{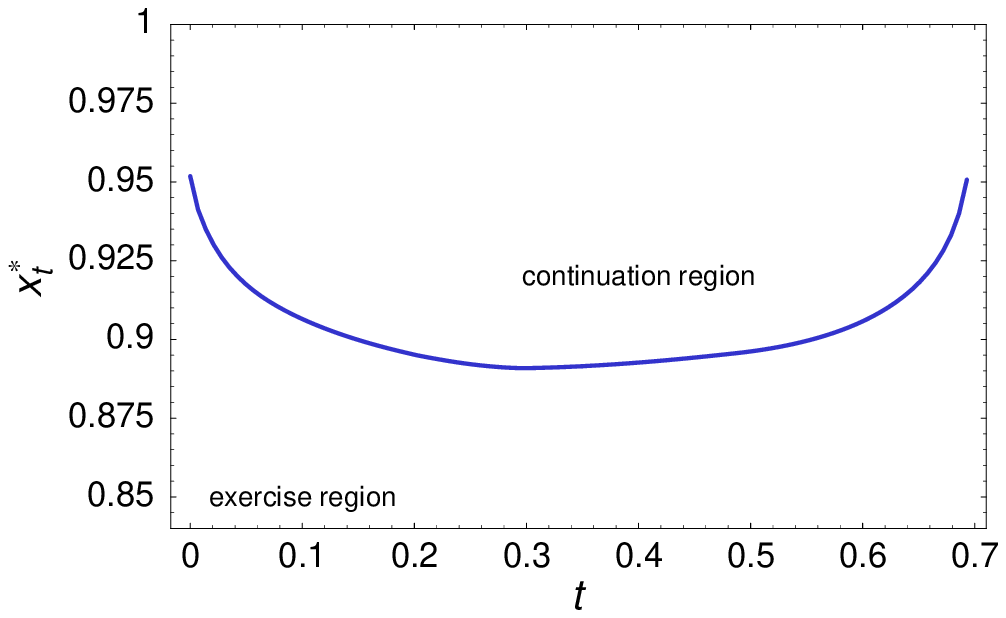}
				\includegraphics[width=0.32\textwidth]{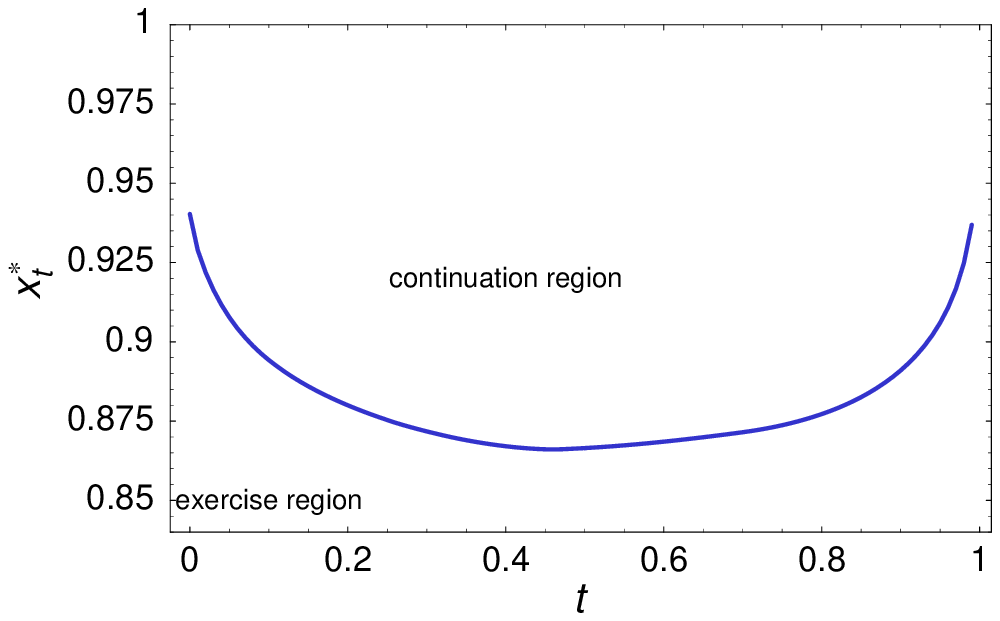}
				\includegraphics[width=0.32\textwidth]{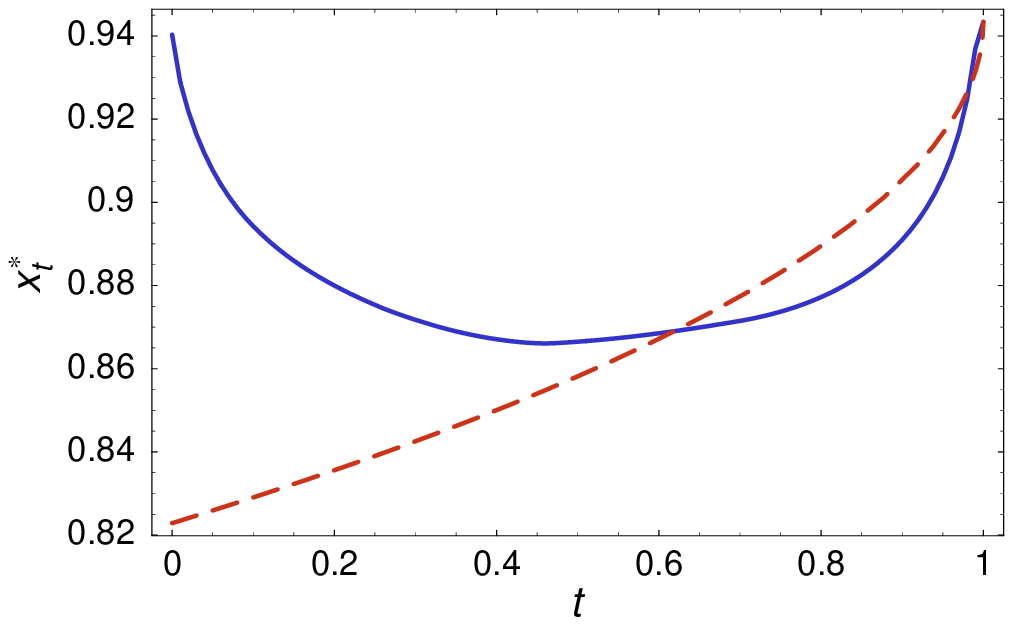}
			\end{center}
			\caption{\small 
			The free boundary position for  expiration times $T=0.7$ (left) and $T=1$  (center). A comparison of the free boundary position
			with its  the analytic approximation (dashed line).
			}
			\label{asian-fig-6}
		\end{figure}

		Finally, in Fig.~\ref{asian-fig-6} we present numerical experiments for shorter expiration times $T=0.7$ and $T=1$ (one year) with zero
		dividend rate $q=0$ and $r=0.06, \sigma=0.2$. We also present a  comparison of the free boundary position $x^*_t=1/\varrho(T-t)$ and
		the analytic approximation (\ref{aa-asymptotika}) for parameters: $r=0.06,q=0,\sigma=0.2$ and $T=1$. It is clear that the analytic
		approximation (\ref{aa-asymptotika}) is capable of capturing the behavior of $x^*_t$ only for times $t$ close to the expiry $T$.
		Moreover, the analytic approximation is a monotone function whereas the true early exercise boundary $x^*_t$ is a decreasing function
		for small values of $t$ and then it becomes increasing (see e.g. Fig.~\ref{asian-fig-6}).

%%%%%%%%%%%%%%%%%%%%%%%%%%%%%%%%%%%%%%%%%%%%%%%%%%%%%%%%%%

\section*{Conclusions}

In this paper we analyzed American-style Asian options with averaged floating strike. We focused on arithmetic, geometric and weighted arithmetic averaging of the floating strike price. In the first part of the paper we derived an integral representation of the call and put option prices and we provided an integral equation for the free boundary position. We analyzed the behavior of the early exercise boundary close to expiry. We proposed a general methodology how to determine the early exercise position at expiry. We furthermore derived the asymptotic formula for the early exercise boundary close to the expiry. The second part of the paper was devoted to construction of a robust numerical scheme for finding an approximation of the early exercise boundary. Applying the front fixing method, we derived a nonlocal parabolic partial differential for the synthetised portfolio and the free boundary position. Using an idea of the operator splitting technique we moreover  constructed a numerical scheme for numerical solution of the problem. The capability of the method has been documented by several computational examples.

\section*{Acknowledgments} This research was supported by VEGA 1/0381/09 and APVV SK-BG-0034-08 grants.

{\small
	\bibliographystyle{kluwer}
	\bibliography{asian-bs-091207}
}
\end{document}